\documentclass[aps,citeautoscript,elide,prb,superscriptaddress,showpacs,twocolumn]{revtex4-1}
\usepackage{mathptmx}
\usepackage[T1]{fontenc}
\usepackage[latin9]{inputenc}
\usepackage{color}
\usepackage{prettyref}
\usepackage{amsmath}
\usepackage{amssymb}
\usepackage{esint}
\usepackage[unicode=true,
 bookmarks=true,bookmarksnumbered=false,bookmarksopen=true,bookmarksopenlevel=3,
 breaklinks=true,pdfborder={0 0 0},backref=false,colorlinks=true]
 {hyperref}
\hypersetup{pdftitle={Current injection by coherent one- and two-photon excitation in graphene and its bilayer},
 pdfauthor={Julien Rioux},
 pdfsubject={graphene; bilayer; current; injection},
 citecolor=blue,linkcolor=blue,pdfpagemode=UseNone,pdfstartview=FitH,urlcolor=blue}
\usepackage{breakurl}

\makeatletter
 
 \@ifundefined{textcolor}{}
 {%
   \definecolor{BLACK}{gray}{0}
   \definecolor{WHITE}{gray}{1}
   \definecolor{RED}{rgb}{1,0,0}
   \definecolor{GREEN}{rgb}{0,1,0}
   \definecolor{BLUE}{rgb}{0,0,1}
   \definecolor{CYAN}{cmyk}{1,0,0,0}
   \definecolor{MAGENTA}{cmyk}{0,1,0,0}
   \definecolor{YELLOW}{cmyk}{0,0,1,0}
 }

\usepackage{hypcap}
\usepackage{svn}
\usepackage{graphicx}
\usepackage{times}
\usepackage{upgreek}

\SVN $Date: 2011-03-02 16:06:20 -0500 (Wed, 02 Mar 2011) $
\SVN $Revision: 262 $

\renewcommand{\vec}[1]{\boldsymbol{\mathbf{#1}}}

\renewcommand{\Im}{\mathbb{I\mathrm{m}}}

\AtBeginDocument{\renewcommand{\eqref}[1]{\hyperref[#1]{(\ref*{#1})}}}
\newrefformat{eq}{\hyperref[#1]{Eq.~(\ref*{#1})}}
\newrefformat{eqs}{Eqs.~\hyperref[#1]{(\ref*{#1})}}
\newrefformat{equation}{\hyperref[#1]{Equation~(\ref*{#1})}}
\newrefformat{fig}{\hyperref[#1]{Fig.~\ref*{#1}}}
\newrefformat{figure}{\hyperref[#1]{Figure~\ref*{#1}}}
\newrefformat{sec}{\hyperref[#1]{Sec.~\ref*{#1}}}
\newrefformat{subfiga}{\hyperref[#1]{Fig.~\ref*{#1}(a)}}
\newrefformat{subfigb}{\hyperref[#1]{Fig.~\ref*{#1}(b)}}
\newrefformat{subfigc}{\hyperref[#1]{Fig.~\ref*{#1}(c)}}
\newrefformat{subfigd}{\hyperref[#1]{Fig.~\ref*{#1}(d)}}
\newrefformat{subfigab}{\hyperref[#1]{Fig.~\ref*{#1}(a,b)}}
\newrefformat{subfigcd}{\hyperref[#1]{Fig.~\ref*{#1}(c,d)}}
\newrefformat{tab}{\hyperref[#1]{Table~\ref*{#1}}}

\makeatother

\begin{document}

\preprint{Revision \SVNRevision}

\title{Current injection by coherent one- and two-photon excitation in graphene
and its bilayer}

\author{J. Rioux}

\affiliation{Department of Physics and Institute for Optical Sciences, University
of Toronto, 60 St.~George Street, Toronto, Ontario, Canada M5S~1A7}

\author{Guido Burkard}

\affiliation{Department of Physics, University of Konstanz, D-78457 Konstanz,
Germany}

\author{J.~E. Sipe}

\affiliation{Department of Physics and Institute for Optical Sciences, University
of Toronto, 60 St.~George Street, Toronto, Ontario, Canada M5S~1A7}

\date{\SVNDate}
\begin{abstract}
Coherent control of optically-injected carrier distributions in single
and bilayer graphene allows the injection of electrical currents.
Using a tight-binding model and Fermi's golden rule, we derive the
carrier and photocurrent densities achieved via interference of the
quantum amplitudes for two-photon absorption at a fundamental frequency,
$\omega$, and one-photon absorption at the second harmonic, $2\omega$.
Strong currents are injected under co-circular and linear polarizations.
In contrast, opposite-circular polarization yields no net current.
For single-layer graphene, the magnitude of the current is unaffected
by the rotation of linear-polarization axes, in contrast with the
bilayer and with conventional semiconductors. The dependence of the
photocurrent on the linear-polarization axes is a clear and measurable
signature of interlayer coupling in AB-stacked multilayer graphene.
We also find that single and bilayer graphene exhibit a strong, distinct
linear-circular dichroism in two-photon absorption.
\end{abstract}



\maketitle

\global\long\def\bra#1{\left\langle #1\right|}
\global\long\def\ket#1{\left|#1\right\rangle }

\global\long\def\cc{\mathrm{c.c.}}
\global\long\def\d{\mathrm{d}}
\global\long\def\Hamiltonian{H}
\global\long\def\Heaviside{\Theta}
\global\long\def\phase{\varphi}
\global\long\def\yogh{d}

\section{Introduction}

The successful isolation of a single graphene sheet \citep{Novoselov2004,Novoselov2005a}
has sparked an intense research area around its unusual electronic
and optical properties. Carriers in graphene obey Dirac's equation,
resulting in an electronic energy-momentum dispersion that is linear,
with intersecting electron and hole bands \citep{Novoselov2005b,Geim2007,CastroNeto2009}.
At optical frequencies, the absorption per layer through a graphene
stack is quantized in an amount written in terms of universal constants
\citep{Nair2008}.

Bilayer graphene has also garnered significant interest due to its
quite different but equivalently interesting electronic properties.
The carriers in clean, unbiased bilayer graphene obey a massive Dirac
equation; their band dispersion is gapless, quadratic at low energy
and linear at high energy \citep{Novoselov2006,Katsnelson2006a}.

Both single and bilayer graphene are characterized by carrier mobilities
that are extremely high \citep{Novoselov2004,Morozov2008a}. Their
high optical conductivity and high carrier mobilities mean they could
see applications as optically-controlled transport devices.

For a level system subjected to coherent irradiation at a fundamental
frequency and its $\ell^{\textit{th}}$ harmonic, the quantum interference
of one- and $\ell$-photon absorption pathways allows the coherent
control (CC) of the excitation process. This quantum interference
control technique has been widely used to study systems ranging from
molecules to bulk and quantum well materials \citep{coherent-control-review}.
For crystalline materials, where initial and final states are described
by Bloch states, an often-studied method is the use of fundamental
and second-harmonic frequencies. The two equivalent pathways consist
of two-photon absorption of the fundamental and one-photon absorption
of the second harmonic. The cross-term of the transition amplitudes
contributes to an asymmetrical distribution of injected carriers through
reciprocal space, yielding a nonzero current density. The $k$-space
distribution is controlled by attributes of the two coherent components
of the light field: their polarization and a relative phase parameter
\citep{Atanasov1996}. In graphene, the interference effect for linearly-polarized
light has been predicted to be significantly stronger than in conventional
semiconductors \citep{Mele2000}. Photocurrent CC has been demonstrated
experimentally in multilayer epitaxial graphene \citep{Sun2010},
carbon nanotubes, and graphite \citep{Newson2008}.

In this paper, the tight-binding model of graphene near the Dirac
point is used to calculate the distributions of carriers optically
injected by simultaneous irradiation with light at a frequency $\omega$
and light at its second harmonic frequency $2\omega$. We find that
the coherent adjustment of phase parameters and polarizations yield
a photocurrent for linearly and co-circularly polarized light. The
results for one and two layers of graphene are contrasted, and we
argue that interlayer coupling could be probed in a Bernal stack of
graphene sheets.

The paper is organized as follows. The effective Hamiltonians used
for the calculations are presented in \prettyref{sec:Hamiltonians}.
One- and two-photon absorption coefficients in single-layer graphene,
the resulting distributions of injected carriers, and the generated
photocurrents due to interference are presented in \prettyref{sec:Graphene};
these results hold as well for the low-energy expansion of bilayer
graphene. The full treatment of the bilayer and the effects of interlayer
coupling on two-photon absorption and photocurrent injection are presented
in \prettyref{sec:Bilayer}. We summarize and discuss our results
in \prettyref{sec:Summary}.

\section{Hamiltonians\label{sec:Hamiltonians}}

\begin{figure}
\capstart

\includegraphics{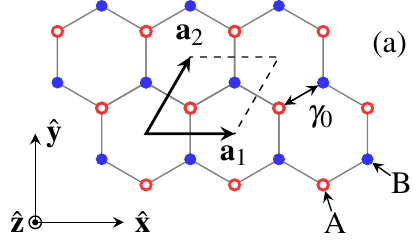}
\quad
%
\includegraphics{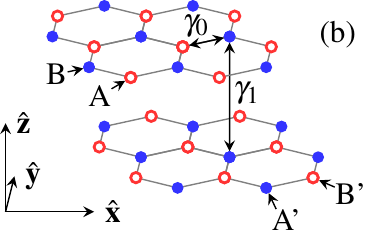}

\caption{Crystal structure of (a) single-layer graphene and (b) bilayer graphene.
The basis vectors $\vec{a}_{1}$ and $\vec{a}_{2}$ define the unit
cell, $\gamma_{0}$ and $\gamma_{1}$ are the intralayer and interlayer
coupling strengths, and the sublattices are denoted by A (A') and
B (B'). \label{fig:crystal-structures} \label{subfiga:graphene-crystal-structure}
\label{subfigb:bilayer-crystal-structure}}
\end{figure}
\begin{figure}
\capstart

\includegraphics{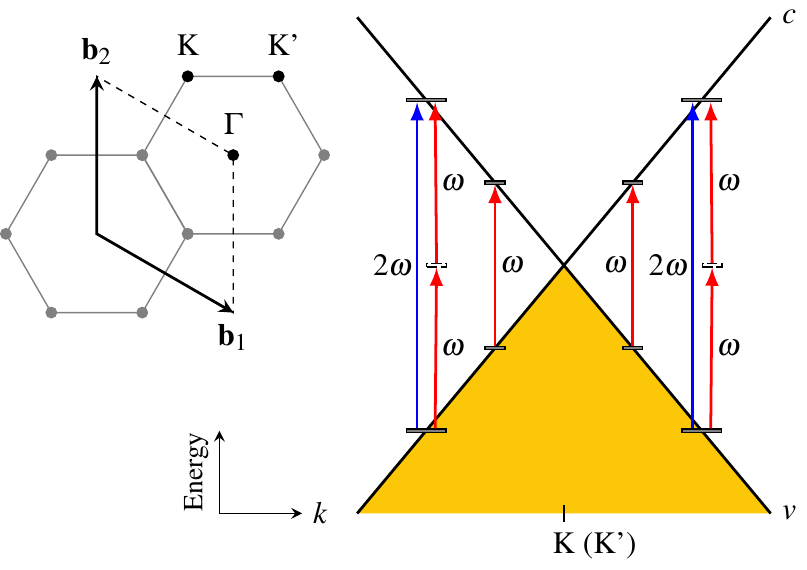}

\caption{(Color online) Reciprocal space and linear energy-crystal momentum
dispersion of graphene near K. The basis vectors $\vec{b}_{1}$ and
$\vec{b}_{2}$ form the reciprocal unit cell, enclosing one K and
one K' valley. The dispersion shows the initially empty conduction
band $c$ and occupied valence band $v$ touching at the K~point.
The excitation scheme employs interference between two-photon absorption
at $\omega$ (red arrows) and one-photon absorption at $2\omega$
(blue arrows), leading to generation of charge and current. \label{fig:reciprocal-space}
\label{fig:graphene-band-structure}}
\end{figure}

Single-layer graphene (henceforth simply graphene) is a one-atom-thick
layer of carbon atoms arranged in two triangular sublattices \{A,
B\}, as shown in \prettyref{subfiga:graphene-crystal-structure}.
In the basis of the sublattices, the tight-binding model is expanded
near the K~point to yield the effective Hamiltonian\begin{align}
\Hamiltonian_{1}(\vec{K}+\vec{k}) & \rightarrow\hbar v_{F}\vec{\sigma}\cdot\vec{k}=v_{F}\left(\begin{array}{cc}
0 & \hbar k_{-}\\
\hbar k_{+} & 0\end{array}\right),\label{eq:graphene-Hamiltonian}\end{align}
where $v_{F}$ is the Fermi velocity, $\vec{\sigma}$ are the Pauli
matrices, $\vec{k}$ is the crystal momentum in the plane of the crystal
relative to the K~point and $k_{\pm}=k_{x}\pm ik_{y}$. The Fermi
velocity can be expressed in terms of the sublattice hopping term
$\gamma_{0}$ \citep{Geim2007,CastroNeto2009}. The resulting band
energies are linear in crystal momentum $k$ and are shown in \prettyref{fig:graphene-band-structure}.

The structure of Bernal-stacked bilayer graphene is sketched in \prettyref{subfigb:bilayer-crystal-structure};
there are four atoms per unit cell, each contributing a $p_{z}$ orbital
to the $\pi$~bands \citep{Nicol2008}. The corresponding $4\times4$
tight-binding Hamiltonian, written in the basis \{A, B', A', B\} and
expanded near the K~point, is given by\begin{equation}
\Hamiltonian_{2}(\vec{K}+\vec{k})\rightarrow\left(\begin{array}{cccc}
0 & 0 & 0 & v_{F}\hbar k_{-}\\
0 & 0 & v_{F}\hbar k_{+} & 0\\
0 & v_{F}\hbar k_{-} & 0 & \gamma_{1}\\
v_{F}\hbar k_{+} & 0 & \gamma_{1} & 0\end{array}\right).\label{eq:bilayer-Hamiltonian}\end{equation}
This tight-binding model includes $\gamma_{0}$, the intralayer coupling,
and $\gamma_{1}$, the hopping term between sublattices A' and B from
the two different layers. A low-energy expansion yields\begin{equation}
\Hamiltonian_{2}^{\prime}(\vec{K}+\vec{k})\rightarrow-\frac{\hbar^{2}}{2m}\left(\begin{array}{cc}
0 & k_{-}^{2}\\
k_{+}^{2} & 0\end{array}\right),\label{eq:bilayer-chiral-Hamiltonian}\end{equation}
where $m=\gamma_{1}/2v_{F}^{2}$. This is the {}``massive'' Dirac
equation, describing the carriers near the K~point for $\left|v_{F}\hbar k\right|\ll\gamma_{1}$.
The energy dispersion consists of a pair of gapless conduction and
valence bands touching at the Dirac point $(\vec{k}=0)$, with a quadratic
dependence on crystal momentum \citep{Geim2007,CastroNeto2009}.

Near the K'~point, similar Hamiltonians are obtained by letting $k_{x}\rightarrow-k_{x}$
in Eqs.~(\ref{eq:graphene-Hamiltonian}--\ref{eq:bilayer-chiral-Hamiltonian});
for the purposes of this paper, the two valleys are equivalent.

\section{Single-layer graphene\label{sec:Graphene}}

In this section, we use \prettyref{eq:graphene-Hamiltonian} to calculate
one- and two-photon absorption coefficients, and the CC of chiral
carriers in graphene. The velocity operator $\vec{v}=\frac{1}{\hbar}\nabla_{\vec{k}}\Hamiltonian$,
when written in the eigenstates basis, takes the form\begin{equation}
\vec{v}\rightarrow v_{F}\left(\begin{array}{cc}
\hat{k} & i\hat{\phi}\\
-i\hat{\phi} & -\hat{k}\end{array}\right),\label{eq:velocity-operator-in-graphene}\end{equation}
where $\hat{k}$ is the unit vector parallel to the direction of $\vec{k}$
and $\hat{\phi}=\hat{z}\times\hat{k}$.

\subsection{Carrier injection}

We calculate the rate of change of the carrier density due to an interaction
Hamiltonian $\Hamiltonian_{\text{int}}=-\frac{e}{m_{0}c}\vec{A}\cdot\vec{p}$,
where $\vec{A}$ is the vector potential of the optical field, $m_{0}$
is the free-electron mass, $c$ is the speed of light in vacuum and
$e=-\left|e\right|$ is the electron charge, by performing a perturbation
calculation up to second order. Assuming a monochromatic field of
frequency $\omega$, we obtain expressions for the rate of injection
of carrier density due to one- and two-photon absorption processes
(using the Gaussian system of quantities and \emph{cgs} units throughout):{\allowdisplaybreaks\begin{align}
\dot{n}_{1} & =\xi_{1}^{ab}(\omega)E^{a*}(\omega)E^{b}(\omega),\label{eq:ndot1}\\
\dot{n}_{2} & =\xi_{2}^{abcd}(\omega)E^{a*}(\omega)E^{b*}(\omega)E^{c}(\omega)E^{d}(\omega),\label{eq:ndot2}\end{align}
}where $\vec{E}$ is the electric field and superscripts $a$, $b$,
$c$, and $d$ indicate Cartesian components; repeated superscripts
are summed over. Microscopic expressions for the tensors $\xi_{1}$
and $\xi_{2}$ are derived in the independent-particle approximation
following Fermi's golden rule (FGR) \citep{Atanasov1996,VanDriel2000}.
For a two-dimensional crystal, we have\begin{multline}
\dot{n}_{\ell}=2\pi\sum_{c,v}\int\frac{\d^{2}k}{4\pi^{2}}\,\left|\Omega_{cv}^{(\ell)}(\omega,\vec{k})\right|^{2}\,\delta[\omega_{cv}(\vec{k})-\ell\omega],\label{eq:ndot-ell}\end{multline}
where $\omega_{cv}(\vec{k})\equiv\omega_{c}(\vec{k})-\omega_{v}(\vec{k})$,
$\hbar\omega_{m}(\vec{k})$ are the band energies, and $\Omega_{cv}^{(\ell)}(\omega,\vec{k})$
is the $\ell$-photon transition amplitude between valence band $v$
and conduction band $c$ at wavevector $\vec{k}$ \citep{VanDriel2000}:
{\allowdisplaybreaks\begin{align}
\Omega_{cv}^{(1)}(\omega,\vec{k}) & =\frac{ie}{\hbar\omega}\vec{v}_{cv}(\vec{k})\cdot\vec{E}(\omega),\label{eq:one-photon-transition-amplitude}\\
\Omega_{cv}^{(2)}(\omega,\vec{k}) & =\frac{2e^{2}}{\hbar^{2}\omega^{2}}\sum_{m}\frac{\vec{v}_{cm}(\vec{k})\cdot\vec{E}(\omega)\:\vec{v}_{mv}(\vec{k})\cdot\vec{E}(\omega)}{\omega_{mc}(\vec{k})+\omega_{mv}(\vec{k})},\label{eq:two-photon-transition-amplitude}\end{align}
}where $\vec{v}_{mn}(\vec{k})$ indicate matrix elements of the velocity
operator, and $[\omega_{mc}(\vec{k})+\omega_{mv}(\vec{k})]/2=\omega_{m}(\vec{k})-[\omega_{v}(\vec{k})+\omega]$
is the usual energy denominator appearing in second-order perturbation
theory.

By the symmetry of graphene and bilayer graphene, the tensors $\xi_{1}$
and $\xi_{2}$ have respectively one and three nonzero independent
components in the $xy$ plane: $\xi_{1}^{xx}$, $\xi_{2}^{xxxx}$,
$\xi_{2}^{xxyy}$, and $\xi_{2}^{xyxy}=\xi_{2}^{xyyx}$; however,
all our model Hamiltonians are isotropic, reducing $\xi_{2}$ to two
independent terms: $\xi_{2}^{xxxx}$ and the linear-circular dichroism
$\delta=\xi_{2}^{xxyy}/\xi_{2}^{xxxx}$ \citep{note-on-isotropic-model}.

The electric field $\vec{E}(\omega)$ in an arbitrary beam at normal
incidence can be written as $\vec{E}(\omega)=E_{\omega}e^{i\phase_{\omega}}(\hat{\vec{x}}_{\omega}+\hat{\vec{y}}_{\omega}e^{i\delta\phase_{\omega}})/\sqrt{2}$,
for an appropriate choice of orthonormal vectors $\hat{\vec{x}}_{\omega}$
and $\hat{\vec{y}}_{\omega}$ in the $xy$ plane, a real amplitude
$E_{\omega}$, and real phase parameters $\phase_{\omega}$ and $\delta\phase_{\omega}$.
The injection rates of the carrier density due to one- and two-photon
processes are given by{\allowdisplaybreaks\begin{align}
\dot{n}_{1} & =\xi_{1}^{xx}(\omega)\left|E_{\omega}\right|^{2},\label{eq:ndot1-in-isotropic-media}\\
\dot{n}_{2} & =\xi_{2}^{xxxx}(\omega)\left|E_{\omega}\right|^{4}\left(1-\delta\sin^{2}(\delta\phase_{\omega})\right).\label{eq:ndot2-in-isotropic-media}\end{align}
}Both are insensitive to rotation of the crystal axes with respect
to the normal, but one-photon absorption is independent of polarization,
while two-photon absorption depends on the phase difference $\delta\phase_{\omega}$
between the linearly-polarized components of the incident light.

For the linear response, we find from \prettyref{eq:graphene-Hamiltonian}
that $\xi_{1}^{xx}\equiv\bar{\xi}_{1}$, with \begin{equation}
\bar{\xi}_{1}(\omega)=2\sigma_{0}/\hbar\omega,\label{eq:xi1-bar-graphene}\end{equation}
where $\sigma_{0}$ is the universal optical conductivity of graphene:
$\sigma_{0}=g_{s}g_{v}\frac{e^{2}}{16\hbar}$, with $g_{s}=2$ and
$g_{v}=2$ denoting spin and valley degeneracy, respectively \citep{Gusynin2006b,Nair2008}.
For the two-photon process we find $\xi_{2}^{xxxx}=\xi_{2}^{xyxy}=\xi_{2}^{xyyx}=-\xi_{2}^{xxyy}\equiv\bar{\xi}_{2}$,
with\begin{equation}
\bar{\xi}_{2}(\omega)=8g_{s}g_{v}\hbar e^{4}v_{F}^{2}\left(2\hbar\omega\right)^{-5}.\label{eq:xi2-bar-graphene}\end{equation}
Thus, for chiral carriers, $\delta=-1$ and it follows from \prettyref{eq:ndot2-in-isotropic-media}
that circularly-polarized light ($\delta\phase_{\omega}=\pm\frac{\pi}{2}$)
provides twice as much two-photon absorption as linearly-polarized
light.

\subsection{Quantum interference of fundamental\protect \\
and second harmonic components}

In the presence of a two-color optical field with frequency components
$\omega$ and $2\omega$, there exist two transition amplitudes connecting
the same initial and final states: $\Omega_{cv}^{(1)}(2\omega,\vec{k})$
results from light at $2\omega$ to first order in perturbation, and
$\Omega_{cv}^{(2)}(\omega,\vec{k})$ results from light at $\omega$
to second order in perturbation. The cross-term of these amplitudes
yields the CC term. Although this has no effect on the total number
of carriers optically injected in centrosymmetric crystals,%
\cite{Note1}
\thanks{Looking at how the interference term in the carrier injection transforms
under the action of inversion symmetry, one gets $\dot{n}_{I}=\xi_{I}^{abc}(\omega)E^{a*}(\omega)E^{b*}(\omega)E^{c}(2\omega)+\cc\rightarrow\dot{n}_{I}=-\xi_{I}^{abc}(\omega)E^{a*}(\omega)E^{b*}(\omega)E^{c}(2\omega)+\cc$,
which imposes $\xi_{I}=0$.%
} it yields an injection term for the current density. This term has
the form\begin{equation}
\dot{J}^{a}=\eta_{I}^{abcd}(\omega)E^{b*}(\omega)E^{c*}(\omega)E^{d}(2\omega)+\cc,\label{eq:Jdot}\end{equation}
where $\eta_{I}(\omega)$ is a fourth-rank current-injection tensor
\citep{Atanasov1996}. The symmetry of graphene or bilayer graphene
yields $\eta_{I}^{xxxx}$, $\eta_{I}^{xyyx}$, and $\eta_{I}^{xyxy}=\eta_{I}^{xxyy}$
as independent components; an isotropic model has $2\eta_{I}^{xyxy}=\eta_{I}^{xxxx}-\eta_{I}^{xyyx}$
\citep{Najmaie2003,Bhat2006}. We introduce a disparity parameter
$\yogh=\eta_{I}^{xyyx}/\eta_{I}^{xxxx}$ to characterize how the current
injection due to linearly-polarized beams depends on whether the polarization
axes are perpendicular or parallel:\begin{multline}
\dot{\vec{J}}=\eta_{I}^{xxxx}(\omega)\Bigl(\vec{E}^{*}(\omega)\left[\vec{E}^{*}(\omega)\cdot\vec{E}(2\omega)\right]\\
-\yogh\,\vec{E}^{*}(\omega)\times\left[\vec{E}^{*}(\omega)\times\vec{E}(2\omega)\right]\Bigr)+\cc\label{eq:Jdot-in-isotropic-media}\end{multline}

From \prettyref{eq:graphene-Hamiltonian} we find that the nonzero
components of the current-injection tensor are related by $\eta_{I}^{xxxx}=\eta_{I}^{xyxy}=\eta_{I}^{xxyy}=-\eta_{I}^{xyyx}\equiv i\bar{\eta}_{I}$,
and thus $\yogh=-1$. In the independent-particle approximation, $\bar{\eta}_{I}$
is purely real. An FGR derivation predicts equal conduction- and valence-band
contributions, for a total current injection\begin{equation}
\bar{\eta}_{I}(\omega)=g_{s}g_{v}e^{4}v_{F}^{2}\left(2\hbar\omega\right)^{-3}.\label{eq:eta-bar-graphene}\end{equation}

We describe $\omega$ and $2\omega$ beams at normal incidence by
the choice of fields $\vec{E}(\omega)=E_{\omega}e^{i\phase_{\omega}}\hat{\vec{e}}_{\omega}$
and $\vec{E}(2\omega)=E_{2\omega}e^{i\phase_{2\omega}}\hat{\vec{e}}_{2\omega}$,
where $\hat{\vec{e}}_{\omega/2\omega}=(\hat{\vec{x}}_{\omega/2\omega}+\hat{\vec{y}}_{\omega/2\omega}e^{i\delta\phase_{\omega/2\omega}})/\sqrt{2}$,
describing two arbitrary normal-incidence beams. We find that co-circularly
polarized beams ($\delta\phase_{\omega}=\delta\phase_{2\omega}=\pm\frac{\pi}{2}$)
yield the current injection with the largest magnitude:\begin{equation}
\dot{\vec{J}}=2\sqrt{2}\bar{\eta}_{I}(\omega)E_{\omega}^{2}E_{2\omega}\,\hat{\vec{m}},\label{eq:Jdot-circular-polarization-in-graphene}\end{equation}
where $\hat{\vec{m}}=\hat{\vec{x}}_{2\omega}\sin(\Delta\phase\mp2\theta)\pm\hat{\vec{y}}_{2\omega}\cos(\Delta\phase\mp2\theta)$.
The phase-difference parameter $\Delta\phase\equiv2\phase_{\omega}-\phase_{2\omega}$
controls the direction of the current; $\theta$ is the angle that
separates the polarization axes of the fundamental from those of the
second harmonic: $\hat{\vec{x}}_{\omega}=\hat{\vec{x}}_{2\omega}\cos\theta+\hat{\vec{y}}_{2\omega}\sin\theta$.
Opposite-circular polarizations ($-\delta\phase_{\omega}=\delta\phase_{2\omega}=\pm\frac{\pi}{2}$)
yield no net current injection, while linearly-polarized beams ($\delta\phase_{\omega}=\delta\phase_{2\omega}=0$)
yield\begin{equation}
\dot{\vec{J}}=2\bar{\eta}_{I}(\omega)E_{\omega}^{2}E_{2\omega}\sin(\Delta\phase)\,\hat{\vec{n}},\label{eq:Jdot-linear-polarization-in-graphene}\end{equation}
where $\hat{\vec{n}}=\hat{\vec{e}}_{2\omega}\cos(2\theta)+\hat{\vec{e}}_{2\omega}^{\perp}\sin(2\theta)$
and $\hat{\vec{e}}_{2\omega}^{\perp}=\hat{\vec{z}}\times\hat{\vec{e}}_{2\omega}$.
Here the angle $\theta$ between polarization axes controls the orientation
of the current within the graphene plane, and $\Delta\phase$ controls
its magnitude. For co-linearly polarized and cross-polarized beams,
the injected current is parallel with the direction of $\hat{\vec{e}}_{2\omega}$,
the polarization axis of the second harmonic. Conversely the injected
current is perpendicular to $\hat{\vec{e}}_{2\omega}$ when the polarization
axes form an angle of $\theta=\frac{\pi}{4}$. Within this model,
the orientation of the crystal axes has no influence on the current
injection at normal incidence.

Graphene seems to be the first material that has been studied for
which any value of $\theta$ is equally effective at injecting a current.
This is in contrast to materials such as GaAs, where one finds $|\eta_{I}^{xyyx}|\ll|\eta_{I}^{xxxx}|$
and thus a configuration with perpendicular polarization axes results
in a significantly weaker current \citep{Atanasov1996,Najmaie2003,Bhat2006}.

\begin{figure}
\capstart

\includegraphics{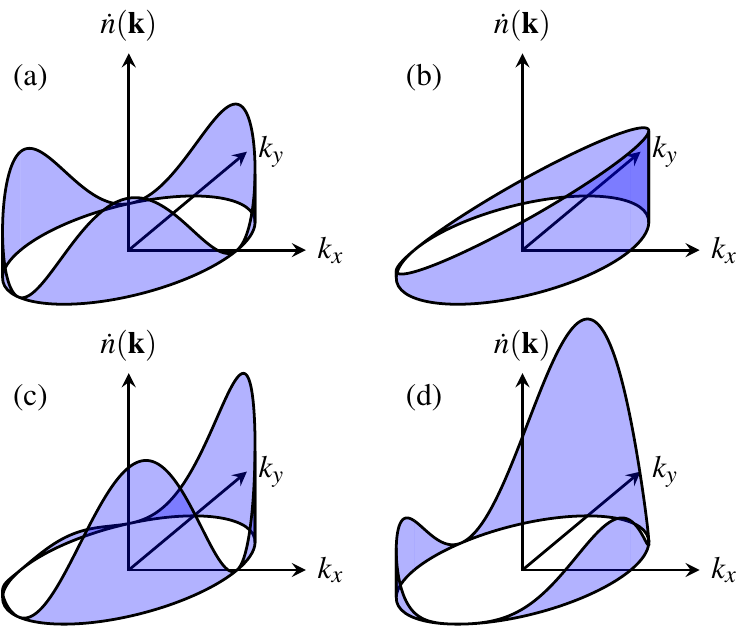}

\caption{Distribution $\dot{n}(\vec{k})$ of the carrier injection through
reciprocal space under irradiation by an optical field with components
$E(\omega)$ and $E(2\omega)$ satisfying $\Delta\phase=\frac{\pi}{2}$.
(a) Opposite-circular polarization ($-\delta\phase_{\omega}=\delta\phase_{2\omega}=\pm\frac{\pi}{2}$,
$\theta=0$). (b) Co-circular polarization ($\sigma^{\pm}$ light,
$\delta\phase_{\omega}=\delta\phase_{2\omega}=\pm\frac{\pi}{2}$,
$\theta=0$). (c,d) Linear polarization ($\delta\phase_{\omega}=\delta\phase_{2\omega}=0$)
with $\hat{\vec{e}}_{2\omega}=\hat{\vec{x}}$ and $\hat{\vec{e}}_{\omega}=\hat{\vec{x}}\cos\theta+\hat{\vec{y}}\sin\theta$;
(c) $\theta=0$ and (d) $\theta=\frac{\pi}{4}$. The distribution
in (a) result in no net current; the asymmetric distributions (b--d)
result in net electrical currents injected in the graphene plane along
$\hat{\vec{x}}$ (b,c) or $\hat{\vec{y}}$~(d). \label{figure:graphene-distributions}
\label{subfiga:graphene-distribution} \label{subfigb:graphene-distribution}
\label{subfigc:graphene-distribution} \label{subfigd:graphene-distribution}
\label{subfigcd:graphene-distribution}}
\end{figure}

\prettyref{figure:graphene-distributions} shows the $k$-space distribution
of the carrier-injection rate, $\dot{n}(\vec{k})=|\Omega_{cv}^{(1)}(2\omega,\vec{k})+\Omega_{cv}^{(2)}(\omega,\vec{k})|^{2}$,
at $\omega_{cv}(\vec{k})=2\omega$. Field amplitudes are chosen such
that the integrated injection rates from one- and two-photon processes
are balanced: $\dot{n}_{1}(2\omega)=\dot{n}_{2}(\omega)$. Opposite-circular
polarization of the beams yield the nonpolar distribution in \prettyref{subfiga:graphene-distribution}
and no net current. In \prettyref{subfigb:graphene-distribution}
both components of the two-color field have the same circular polarization
$\sigma^{\pm}$. The carrier distribution follows $\dot{n}(\vec{k})\varpropto1+\sin(\Delta\phase\pm\phi_{k})$,
where $\phi_{k}=\tan^{-1}(k_{y}/k_{x})$, resulting in the injection
of the current given by \prettyref{eq:Jdot-circular-polarization-in-graphene}.
The charge distribution and current rotate with $\Delta\phase$: clockwise
for $\sigma^{+}$ and counterclockwise for $\sigma^{-}$, when viewed
from $z>0$.

In \prettyref{subfigcd:graphene-distribution} we show the $k$-space
distribution of the carrier-injection rate for linearly-polarized
light; without loss of generality, $\hat{\vec{e}}_{2\omega}$ is taken
along the $x$ axis: $\hat{\vec{e}}_{2\omega}=\hat{\vec{x}}$ and
$\hat{\vec{e}}_{\omega}=\hat{\vec{x}}\cos\theta+\hat{\vec{y}}\sin\theta$.
Taking the phase-difference parameter to be $\Delta\phase=\frac{\pi}{2}$,
we maximize both the cross-term in the $k$-dependent carrier density,
$\dot{n}(\vec{k})\varpropto|\sin\phi_{k}+ie^{-i\Delta\phase}\sin(2\phi_{k}-2\theta)|^{2}$,
and the resulting current, \prettyref{eq:Jdot-linear-polarization-in-graphene}.
For co-linear polarization axes as in \prettyref{subfigc:graphene-distribution},
the distribution is symmetric with respect to $k_{y}$, while asymmetric
and strongly enhanced towards positive $k_{x}$, although with a node
at $\phi_{k}=0$. The excess of positive-$k_{x}$ carriers gives rise
to a net electric current along the $x$ axis. As the polarization
axis of the $\omega$ component is rotated by the angle $\theta$,
carriers are redistributed towards positive $k_{y}$. At $\theta=\frac{\pi}{4}$,
$\dot{n}(\vec{k})$ is symmetric with respect to $k_{x}$ and the
net current is along the $y$ axis {[}\prettyref{subfigd:graphene-distribution}{]}.

\section{Bilayer graphene\label{sec:Bilayer}}

The chiral Hamiltonian for bilayer graphene, \prettyref{eq:bilayer-chiral-Hamiltonian},
results in the same carrier and current injection as in the previous
section if we replace $\bar{\xi}_{1}\rightarrow2\bar{\xi}_{1}$, $\bar{\xi}_{2}\rightarrow8\hbar\omega\bar{\xi}_{2}/\gamma_{1}$,
and $\bar{\eta}_{I}\rightarrow8\hbar\omega\bar{\eta}_{I}/\gamma_{1}$;
the velocity operator in the eigenstates basis takes the form of \prettyref{eq:velocity-operator-in-graphene}
with $v_{F}\rightarrow\hbar k/m$, and thus the symmetry properties
of the injection tensors are unchanged. But such a treatment describes
the carriers only near the K~point for $\left|v_{F}\hbar k\right|\ll\gamma_{1}$
and leaves out important remote bands in the two-photon transition
amplitude. A more accurate model is given by \prettyref{eq:bilayer-Hamiltonian},
which is also valid for band energies on the order of $\gamma_{1}$.
This $4\times4$ Hamiltonian introduces two additional bands, one
above and one below the Dirac point, shifted by an energy $\gamma_{1}$.
More importantly, it gives the correct linear dispersion for larger
values of $k$. The band dispersion near K is shown in \prettyref{fig:bilayer-band-structure}.

\subsection{Carrier injection}

Starting from \prettyref{eq:bilayer-Hamiltonian}, in this section
we repeat the previous injection-tensor calculations for the unbiased
Bernal-stacked graphene bilayer. The one-photon carrier injection
of bilayer graphene is obtained from \prettyref{eq:xi1-bar-graphene}
by replacing $\sigma_{0}$ with the bilayer optical conductivity $\sigma$
from Abergel and Fal'ko \citep{Abergel2007}.

\begin{figure}
\capstart

\includegraphics{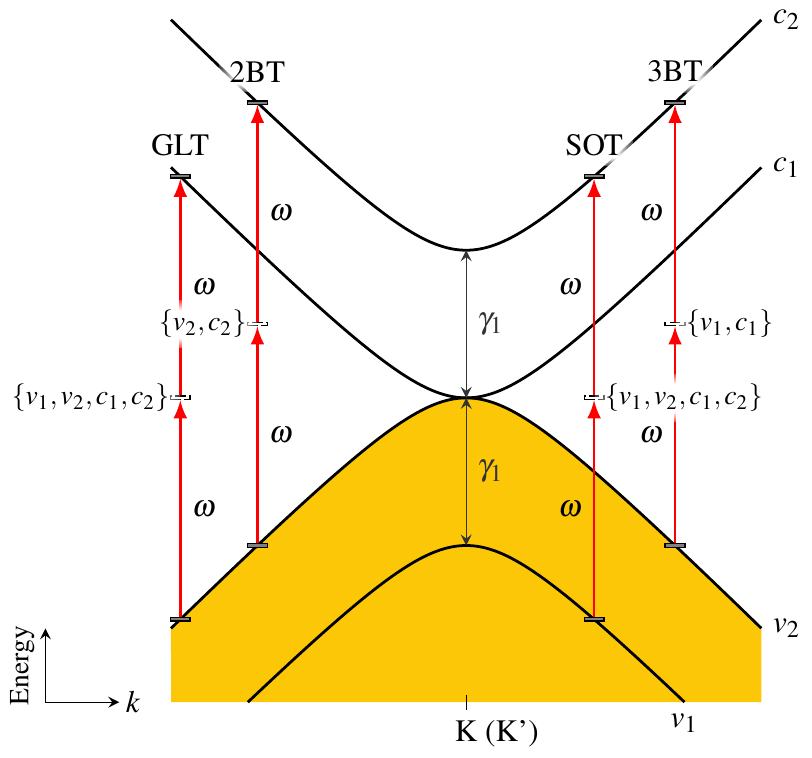}

\caption{(Color online) Band dispersion of bilayer graphene and breakdown of
the transition amplitudes for two-photon absorption. Bands $v_{1}$
and $v_{2}$ are valence bands and initially filled, $c_{1}$ and
$c_{2}$ are initially empty conduction bands. Bands $v_{2}$ and
$c_{1}$ form a gapless doublet touching at the K~point; $c_{2}$
and $v_{1}$ are split-off bands shifted by an energy $\gamma_{1}$
above and below the gapless doublet, respectively. All bands are quadratic
near K and linear at larger $k$. Transition amplitudes appear in
four variants: i) the gapless term (GLT) between bands $v_{2}$ and
$c_{1}$, ii) two- and iii) three-band terms involving exactly one
split-off band (2BT and 3BT, respectively), and iv) the split-off
term (SOT) between bands $v_{1}$ and $c_{2}$. The notation $\{...\}$
next to a virtual state indicates that the sum in \prettyref{eq:two-photon-transition-amplitude}
is restricted to $m\in\{...\}$. Not shown are the 2BT and 3BT between
bands $v_{1}$ and $c_{1}$. \label{fig:bilayer-band-structure} \label{fig:two-photon-contributions-diagrams}}
\end{figure}

We break down the two-photon carrier injection into four distinct
contributions. The first (a) comes from absorption by the gapless
doublet {[}leftmost transition in \prettyref{fig:two-photon-contributions-diagrams},
denoted GLT{]}. The second and third contributions arise from injection
involving exactly one split-off band and contain either (b) only two-band
or three-band amplitudes and no cross-term, or (c) cross-terms of
two- and three-band amplitudes; two- and three-band amplitudes are
denoted 2BT and 3BT in \prettyref{fig:two-photon-contributions-diagrams}.
The fourth contribution (d) comes from absorption where initial and
final states are split-off bands {[}SOT in \prettyref{fig:two-photon-contributions-diagrams}{]}.
\begin{subequations}
The nonzero tensor components are obtained from the symmetry of the
matrix elements involved for each contribution, which yields: \label{eq:xi2-bilayer}\begin{gather}
\xi_{2}^{xxxx}(\omega)=\bar{\xi}_{2a}(\omega)+\left[3\bar{\xi}_{2b}(\omega)+\bar{\xi}_{2c}(\omega)\right]\Heaviside(2\hbar\omega-\gamma_{1})\hphantom{-}\nonumber \\
{}+\bar{\xi}_{2d}(\omega)\,\Heaviside(2\hbar\omega-2\gamma_{1}),\displaybreak[0]\\
\xi_{2}^{xxyy}(\omega)=-\bar{\xi}_{2a}(\omega)+\left[\bar{\xi}_{2b}(\omega)+3\bar{\xi}_{2c}(\omega)\right]\Heaviside(2\hbar\omega-\gamma_{1})\nonumber \\
{}-\bar{\xi}_{2d}(\omega)\,\Heaviside(2\hbar\omega-2\gamma_{1}),\end{gather}
where $\Heaviside(x)$ is the Heaviside step function.
\end{subequations}
The matrix elements appearing in $\bar{\xi}_{2a}$ and $\bar{\xi}_{2d}$
have the same symmetry as those appearing in graphene. However, the
contributions involving exactly one split-off band (b,c) break the
graphene result of $\delta=-1$. Indeed, if one defines a partial
linear-circular dichroism $\delta_{i}=\bar{\xi}_{2i}^{xxyy}/\bar{\xi}_{2i}^{xxxx}$
for each contribution $i\in\{a,b,c,d\}$, it follows that $\delta_{a}=-1$,
$\delta_{b}=\tfrac{1}{3}$, $\delta_{c}=3$ and $\delta_{d}=-1$.
The total linear-circular dichroism will depend on the relative strength
of the contributions $\bar{\xi}_{2a\text{--}d}$, which we now address.

In computing $\Omega_{cv}^{(2)}(\omega,\vec{k})$ for the bilayer,
a difficulty arises since it is possible for the energy denominator
inside the sum in \prettyref{eq:two-photon-transition-amplitude}
to become exactly zero. Take for example the top valence band as initial
state $v$ and the second conduction band as final state $c$. When
the intermediate state $m$ is the first conduction band, there exists
a value of $k$ such that $\hbar\omega_{m}(\vec{k})=\frac{1}{2}\gamma_{1}$;
at this $k$ the intermediate state lies precisely in-between the
initial and final states. This leads to a resonance in the calculated
response functions at $\hbar\omega=\gamma_{1}$. To avoid this resonance,
we let $\omega_{m}\rightarrow\omega_{m}+i\Gamma/2\hbar$ in \prettyref{eq:two-photon-transition-amplitude}.
The linewidth $\Gamma$ accounts phenomenologically for dephasing
due to actual population of the intermediate state. Other linewidths
could be added to describe the effects of disorder or interactions,
but for values $\lesssim\Gamma$ we find that their inclusion does
not significantly modify our results.

\begin{figure}
\capstart

\global\long\def\y{0.35}
\global\long\def\ya{0.15}
\global\long\def\yb{0.25}
\global\long\def\yc{0.35}
\global\long\def\yd{0.45}

\includegraphics{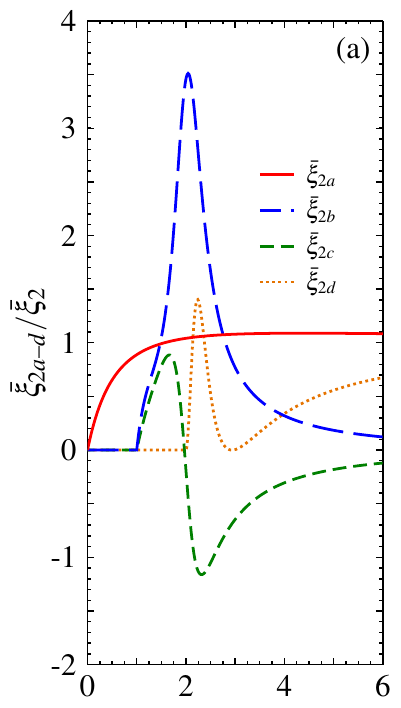}
\includegraphics{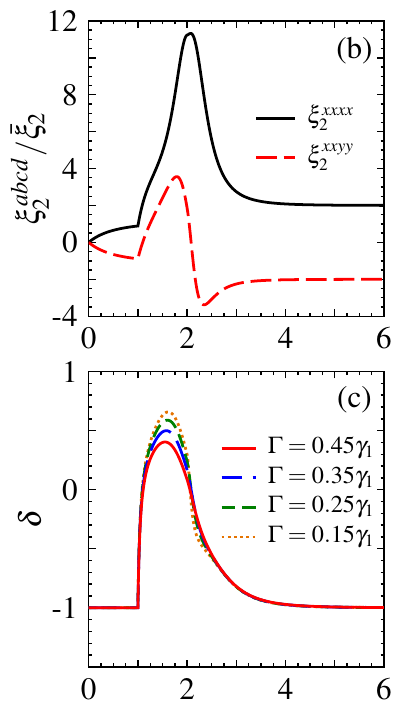}

Photon energy ($2\hbar\omega/\gamma_1$)

\caption{(Color online) Quantities describing two-photon carrier injection
in bilayer graphene as a function of photon energy for an intermediate
state linewidth $\Gamma/\gamma_{1}=\y$. (a) The individual contributions
$\bar{\xi}_{2a\text{--}d}$ ($a$, plain red; $b$, long-dashed blue;
$c$, short-dashed green; $d$, dotted orange) from \prettyref{eq:xi2-bar-bilayer}.
(b) The independent nonzero tensor components $\xi_{2}^{xxxx}$ (plain
black) and $\xi_{2}^{xxyy}$ (dashed red) from \prettyref{eq:xi2-bilayer}.
(c) The linear-circular dichroism $\delta=\xi_{2}^{xxyy}/\xi_{2}^{xxxx}$.
\label{fig:xi2-bilayer} \label{subfiga:xi2-bar-bilayer} \label{subfigb:xi2-abcd-bilayer}
\label{subfigc:delta-bilayer} \label{subfigab:xi2-bilayer}}
\end{figure}

\begin{subequations}
With $\bar{\xi}_{2}(\omega)$ given in \prettyref{eq:xi2-bar-graphene},
we have \label{eq:xi2-bar-bilayer}\begin{align}
\bar{\xi}_{2a}\left(\omega\right) & =\bar{\xi}_{2}(\omega)\frac{2\hbar\omega\left(2\hbar\omega+3\gamma_{1}\right)^{2}}{\left(2\hbar\omega+\gamma_{1}\right)\left(2\hbar\omega+2\gamma_{1}\right)^{2}},\displaybreak[0]\\
\bar{\xi}_{2b}\left(\omega\right) & =\bar{\xi}_{2}(\omega)\frac{2\gamma_{1}^{2}}{\left(2\hbar\omega\right)^{2}}\frac{\left(2\hbar\omega+\gamma_{1}\right)\left(2\hbar\omega-\gamma_{1}\right)}{\left(2\hbar\omega+2\gamma_{1}\right)^{2}}\nonumber \\
 & \times\left(\frac{\left(2\hbar\omega+2\gamma_{1}\right)^{2}}{\left(2\hbar\omega\right)^{2}}+\frac{\left(2\hbar\omega\right)^{2}+\frac{1}{4}\Gamma^{2}}{\left(2\hbar\omega-2\gamma_{1}\right)^{2}+\Gamma^{2}}\right),\displaybreak[0]\\
\bar{\xi}_{2c}\left(\omega\right) & =-\bar{\xi}_{2}(\omega)\frac{2\gamma_{1}^{2}}{\left(2\hbar\omega\right)^{3}}\left(2\hbar\omega+\gamma_{1}\right)\left(2\hbar\omega-\gamma_{1}\right)\nonumber \\
 & \times\left(\frac{1}{2\hbar\omega+2\gamma_{1}}+\frac{2\hbar\omega-2\gamma_{1}}{\left(2\hbar\omega-2\gamma_{1}\right)^{2}+\Gamma^{2}}\right),\displaybreak[0]\\
\bar{\xi}_{2d}\left(\omega\right) & =\bar{\xi}_{2}(\omega)\frac{2\hbar\omega\left(2\hbar\omega-2\gamma_{1}\right)^{2}}{\left(2\hbar\omega-\gamma_{1}\right)^{3}}\nonumber \\
 & \times\left(1-\frac{\gamma_{1}^{2}}{\left(2\hbar\omega-2\gamma_{1}\right)^{2}+\Gamma^{2}}\right)^{2}.\end{align}
We graph the individual contributions $\bar{\xi}_{2a\text{--}d}$
in \prettyref{subfiga:xi2-bar-bilayer} and the resulting $\xi_{2}^{xxxx}$
and $\xi_{2}^{xxyy}$ components of the two-photon carrier-injection
tensor in \prettyref{subfigb:xi2-abcd-bilayer}.
\end{subequations}
In contrast to the linear absorption, where the limits of the bilayer
response function at low and high photon energies gave the graphene
result times a factor of 2 \citep{Abergel2007}, in the two-photon
response function this {}``factor of 2'' rule does not hold at low
photon energy: for $\hbar\omega\ll\gamma_{1}$, we find $\xi_{2\text{(bilayer)}}^{abcd}(\omega)\rightarrow9\hbar\omega\xi_{2\text{(graphene)}}^{abcd}(\omega)/2\gamma_{1}$.
Further, using \prettyref{eq:bilayer-chiral-Hamiltonian} instead
yields $\xi_{2\text{(bilayer)}}^{abcd}(\omega)\rightarrow8\hbar\omega\xi_{2\text{(graphene)}}^{abcd}(\omega)/\gamma_{1}$;
the discrepancy is explained since the derivation using the $4\times4$
Hamiltonian includes important three-band terms from the split-off
bands in the two-photon transition amplitude. At high photon energy
$\xi_{2\text{(bilayer)}}^{abcd}(\omega)\rightarrow2\xi_{2\text{(graphene)}}^{abcd}(\omega)$
as expected.

Two features are apparent in the response tensor at the thresholds
for absorption into the split-off bands. The first feature is a pronounced
shoulder in both $\xi_{2}^{xxxx}$ and $\xi_{2}^{xxyy}$ at $2\hbar\omega=\gamma_{1}$
due to the onset of absorption involving one split-off band. The second
feature is the resonance which occurs at $2\hbar\omega=2\gamma_{1}$.
There the contributions $\bar{\xi}_{2b}$ and $\bar{\xi}_{2c}$ approximate
the real and imaginary parts of a complex Lorentzian function, and
contribute to a peak in $\xi_{2}^{xxxx}$ and a change of sign in
$\xi_{2}^{xxyy}$.

In \prettyref{subfigab:xi2-bilayer} we have chosen the particular
value $\Gamma=\y\gamma_{1}$ for the linewidth. We note that $\bar{\xi}_{2a}$
is independent of $\Gamma$, but $\bar{\xi}_{2b\text{--}d}$ are not.
However, the graph of these quantities changes quantitatively but
not qualitatively when a different finite value is chosen for $\Gamma$.
We plot the linear-circular dichroism $\delta$ in \prettyref{subfigc:delta-bilayer}
for different values of $\Gamma$. It can be seen that $\Gamma$ has
very little effect on $\delta$. The graphene result ($\delta=-1$)
is reproduced for $2\hbar\omega<\gamma_{1}$ and $2\hbar\omega\gtrsim3\gamma_{1}$.
For mid-frequencies, $\gamma_{1}<2\hbar\omega<2\gamma_{1}$, the dichroism
increases (decreases) sharply at the first (second) split-off band
threshold; $\delta$ changes sign and has a maximum value $\sim0.5$
near $2\hbar\omega=1.5\gamma_{1}$.

\subsection{Current injection}

\begin{figure}
\capstart

\includegraphics{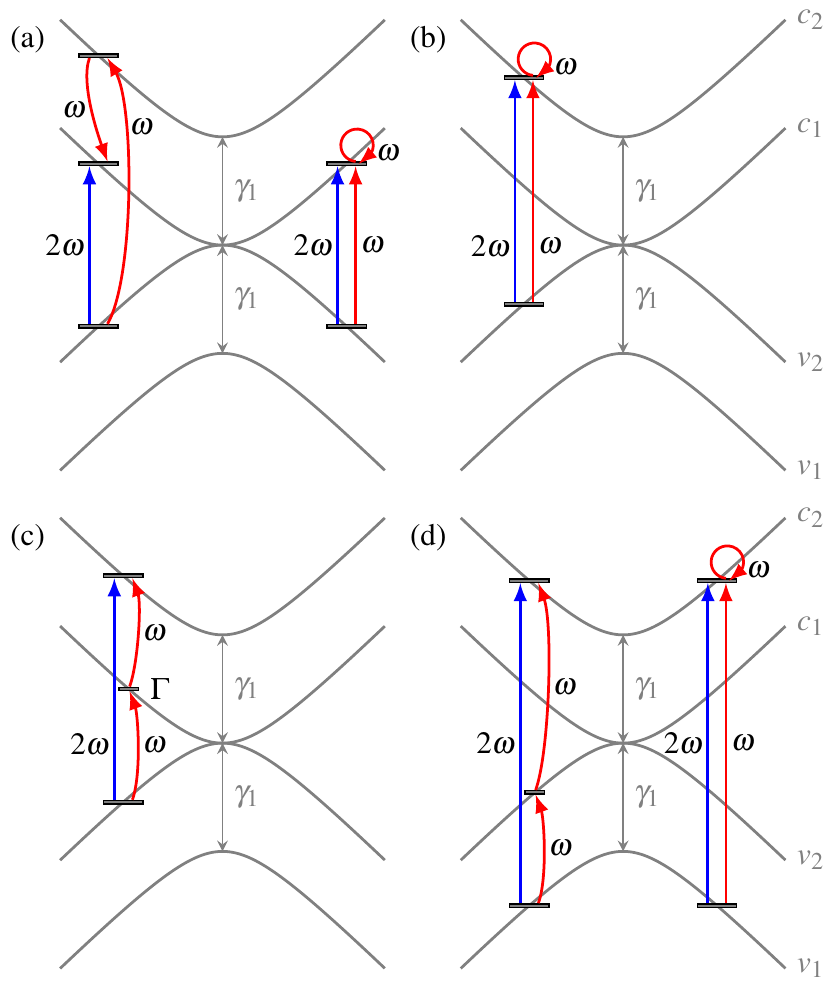}

\caption{Diagrams of the four contributions $\bar{\eta}_{Ia\text{--}d}$ to
the current injection in bilayer graphene {[}\prettyref{eq:eta-bar-bilayer}{]}.\label{fig:current-contributions-diagrams}}
\end{figure}

\begin{figure}
\capstart

\global\long\def\y{0.25}

\includegraphics{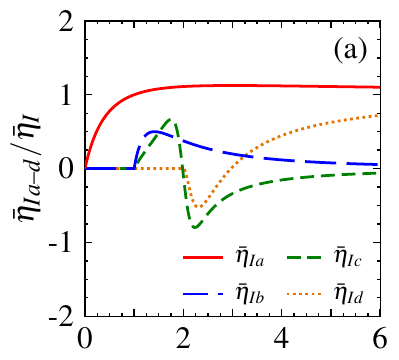}
\includegraphics{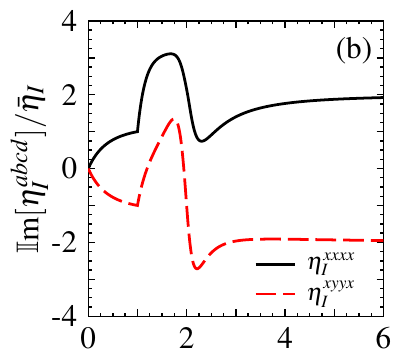}

Photon energy ($2\hbar\omega/\gamma_1$)

\caption{(Color online) The current-injection tensor $\eta_{I}$ in bilayer
graphene for an intermediate state linewidth $\Gamma/\gamma_{1}=\y$.
(a) The four contributions $\bar{\eta}_{Ia\text{--}d}$ from \prettyref{eq:eta-bar-bilayer}:
Plain red, long-dashed blue, short-dashed green, and dotted orange
curves correspond to component $a$, $b$, $c$, and $d$, respectively.
(b) The tensor components $\eta_{I}^{xxxx}$ (plain black) and $\eta_{I}^{xyyx}$
(dashed red). \label{fig:eta-bilayer} \label{subfiga:eta-bar-bilayer}
\label{subfigb:eta-abcd-bilayer}}
\end{figure}

There are four contributions to the current-injection tensor in bilayer
graphene, outlined in \prettyref{fig:current-contributions-diagrams}:
The first (a) comes from absorption by the gapless doublet. The second
and third contributions arise from injection involving exactly one
split-off band, with the second-order amplitude containing either
(b) only two-band terms or (c) only three-band terms. Transitions
involving only the split-off bands make up the fourth contribution
(d). 
\begin{subequations}
Each individual process involves matrix elements of varying symmetry,
and they contribute differently to $\eta_{I}^{xxxx}$ and $\eta_{I}^{xyyx}$:
\label{eq:eta-bilayer}\begin{gather}
\eta_{I}^{xxxx}(\omega)=i\bar{\eta}_{Ia}(\omega)+\left[3i\bar{\eta}_{Ib}(\omega)+i\bar{\eta}_{Ic}(\omega)\right]\Heaviside(2\hbar\omega-\gamma_{1})\hphantom{-}\nonumber \\
+i\bar{\eta}_{Id}(\omega)\,\Heaviside(2\hbar\omega-2\gamma_{1}),\displaybreak[0]\\
\eta_{I}^{xyyx}(\omega)=-i\bar{\eta}_{Ia}(\omega)+\left[i\bar{\eta}_{Ib}(\omega)+3i\bar{\eta}_{Ic}(\omega)\right]\Heaviside(2\hbar\omega-\gamma_{1})\nonumber \\
{}-i\bar{\eta}_{Id}(\omega)\,\Heaviside(2\hbar\omega-2\gamma_{1}).\end{gather}

\end{subequations}
The four contributions have dissimilar values of the parallel-perpendicular
polarization disparity parameter: $\yogh_{a}=-1$, $\yogh_{b}=\frac{1}{3}$,
$\yogh_{c}=3$, and $\yogh_{d}=-1$.
\begin{subequations}
Their magnitudes are given by \label{eq:eta-bar-bilayer}\begin{align}
\bar{\eta}_{Ia}(\omega) & =\bar{\eta}_{I}(\omega)\frac{2\hbar\omega}{\left(2\hbar\omega+\gamma_{1}\right)^{2}}\left(2\hbar\omega+3\gamma_{1}\right),\displaybreak[0]\\
\bar{\eta}_{Ib}(\omega) & =\bar{\eta}_{I}(\omega)\frac{2\gamma_{1}^{2}}{\left(2\hbar\omega\right)^{2}}\frac{\left(2\hbar\omega+\gamma_{1}\right)\left(2\hbar\omega-\gamma_{1}\right)}{\left(2\hbar\omega\right)^{2}},\displaybreak[0]\\
\bar{\eta}_{Ic}(\omega) & =-\bar{\eta}_{I}(\omega)\frac{\gamma_{1}^{2}\left(2\hbar\omega+\gamma_{1}\right)\left(2\hbar\omega-\gamma_{1}\right)}{\left(2\hbar\omega\right)^{3}}\nonumber \\
 & \times\left(\frac{1}{2\hbar\omega+2\gamma_{1}}+\frac{2\hbar\omega-2\gamma_{1}}{\left(2\hbar\omega-2\gamma_{1}\right)^{2}+\Gamma^{2}}\right),\displaybreak[0]\\
\bar{\eta}_{Id}(\omega) & =\bar{\eta}_{I}(\omega)\frac{2\hbar\omega}{\left(2\hbar\omega-\gamma_{1}\right)^{3}}\left(2\hbar\omega-2\gamma_{1}\right)^{2}\nonumber \\
 & \times\left(1-\frac{\gamma_{1}^{2}}{\left(2\hbar\omega-2\gamma_{1}\right)^{2}+\Gamma^{2}}\right),\end{align}
with $\bar{\eta}_{I}(\omega)$ given in \prettyref{eq:eta-bar-graphene}.
\end{subequations}
The parameters $\bar{\eta}_{Ia\text{--}d}$ are plotted in \prettyref{subfiga:eta-bar-bilayer}.
The two independent components of $\eta_{I}$ in the isotropic model,
$\eta_{I}^{xxxx}$ and $\eta_{I}^{xyyx}$, are plotted in \prettyref{subfigb:eta-abcd-bilayer}.
In the high frequency limit, the current-injection tensor for the
bilayer tends to $\eta_{I\text{(bilayer)}}^{abcd}(\omega)\rightarrow2\eta_{I\text{(graphene)}}^{abcd}(\omega)$.
In the low frequency limit, we get $\eta_{I\text{(bilayer)}}^{abcd}(\omega)\rightarrow6\hbar\omega\eta_{I\text{(graphene)}}^{abcd}(\omega)/\gamma_{1}$,
in contrast to using the simple Hamiltonian of \prettyref{eq:bilayer-chiral-Hamiltonian}
which neglects three-band terms in the second-order amplitude and
gave $\eta_{I\text{(bilayer)}}^{abcd}(\omega)\rightarrow8\hbar\omega\eta_{I\text{(graphene)}}^{abcd}(\omega)/\gamma_{1}$
(see the start of \prettyref{sec:Bilayer}).

\begin{figure*}
\capstart

\global\long\def\ya{0.15}
\global\long\def\yb{0.25}
\global\long\def\yc{0.35}
\global\long\def\yd{0.45}

\includegraphics{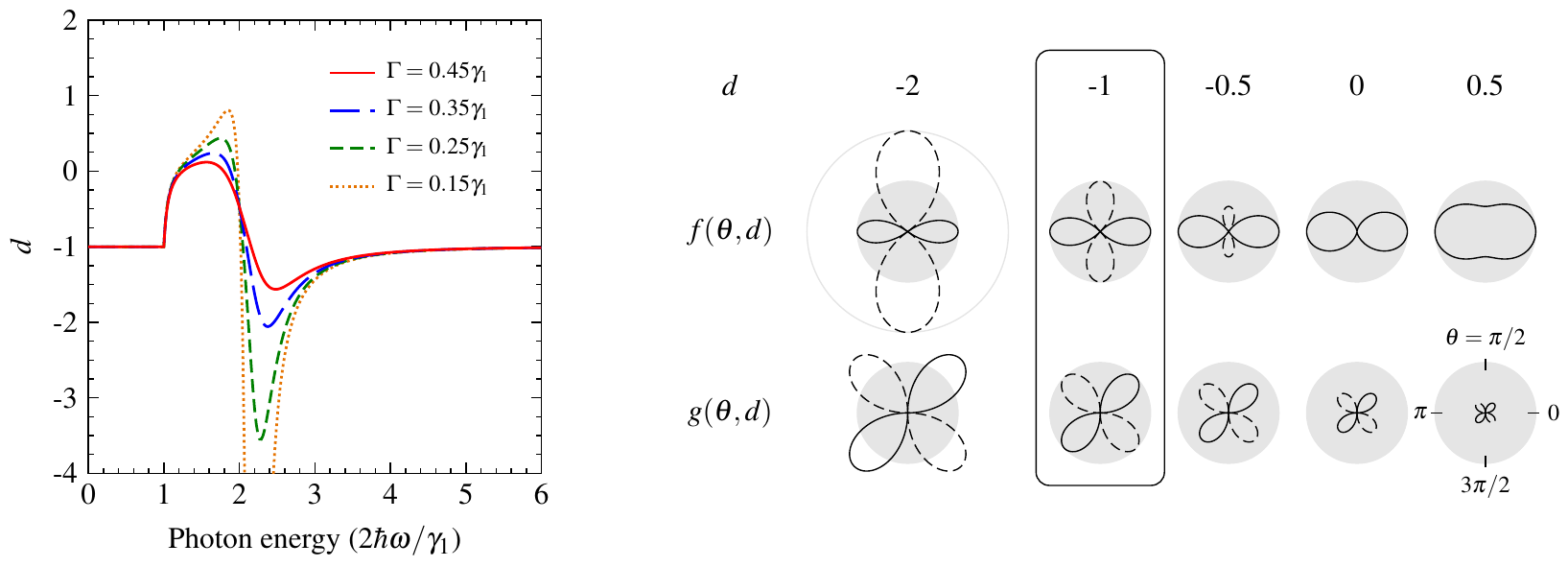}

\caption{(Color online) Results for the current injection in bilayer graphene
with linearly-polarized $\omega$ and $2\omega$ beams. Left-hand
side: The disparity parameter $\yogh=\eta_{I}^{xyyx}/\eta_{I}^{xxxx}$
describing the asymmetry between parallel and perpendicular polarization
axes, with $\Gamma/\gamma_{1}=\ya$ (dotted orange), $\yb$ (short-dashed
green), $\yc$ (long-dashed blue), and $\yd$ (plain red). Right-hand
side: Polar plots of $f(\theta,\yogh)$ and $g(\theta,\yogh)$, the
angular distributions of the projections of $\dot{\vec{J}}$ parallel
and perpendicular to $\hat{\vec{e}}_{2\omega}$, as a function of
the angle $\theta$ between the polarization vectors for $\yogh=-2$,
$-1$, $-0.5$, $0$, and $0.5$. The shaded circles represent unit
amplitude and dashed lines represent negative projections. The graphene
prediction, $\yogh=-1$, is highlighted.\label{fig:yogh-bilayer}}
\end{figure*}

For mid-frequencies, there is a sharp increase in $\eta_{I}$ at the
first split-off band edge at $2\hbar\omega=\gamma_{1}$ and a sharp
decrease at the second edge at $2\hbar\omega=2\gamma_{1}$. Two features
are manifest as a consequence of these split-off band edges: (i) In
the region $\gamma_{1}<2\hbar\omega<2\gamma_{1}$ the $\eta_{I}^{xyyx}$
component changes sign. (ii) For $2\hbar\omega\gtrsim2\gamma_{1}$
the $\eta_{I}^{xxxx}$ component becomes very small.

The main difference between the current injection in graphene and
in the bilayer is the contribution of components $\bar{\eta}_{Ib}$
and $\bar{\eta}_{Ic}$, each with a vastly different value of the
disparity parameter: $\yogh_{b}=\frac{1}{3}$ and $\yogh_{c}=3$ while
in graphene $\yogh=-1$. In \prettyref{fig:yogh-bilayer} we plot
the frequency dependence of $\yogh$ in bilayer graphene. The spectrum
shows a constant $-1$ value from zero frequency until a sharp increase
at the first split-off band edge at $2\hbar\omega=\gamma_{1}$; $\yogh$
rises with photon energy and eventually switches sign. At the second
split-off band edge at $2\hbar\omega=2\gamma_{1}$, $\yogh$ reverses
sign abruptly; for the range $2\gamma_{1}<2\hbar\omega\lesssim3\gamma_{1}$
it takes on large negative values as $\eta_{I}^{xyyx}$ remains finite
but $\eta_{I}^{xxxx}$ becomes small {[}\emph{cf.}~\prettyref{subfigb:eta-abcd-bilayer}{]}.
The value of $\yogh$ tends to $-1$ at higher photon energy.

We now consider current injection in bilayer graphene under irradiation
by $\omega$ and $2\omega$ beams at normal incidence. Choosing the
electric fields $\vec{E}(\omega)$ and $\vec{E}(2\omega)$ as in the
previous section, the current injection is given for co-circular polarization
of the beams ($\delta\phase_{\omega}=\delta\phase_{2\omega}=\pm\frac{\pi}{2}$)
by \begin{equation}
\dot{\vec{J}}=(1-\yogh)\sqrt{2}\Im\left[\eta_{I}^{xxxx}\right]E_{\omega}^{2}E_{2\omega}\hat{\vec{m}}\label{eq:Jdot-co-circular-polarization-in-isotropic-medium}\end{equation}
and for opposite-circular polarization ($-\delta\phase_{\omega}=\delta\phase_{2\omega}=\pm\frac{\pi}{2}$)
by $\dot{\vec{J}}=0$. In \prettyref{eq:Jdot-co-circular-polarization-in-isotropic-medium},
the disparity parameter $d$ only affects the magnitude of the current.
In contrast, for linearly-polarized $\omega$ and $2\omega$ beams
($\delta\phase_{\omega}=\delta\phase_{2\omega}=0$) forming an angle
$\theta$ between their polarization axes, different values of $d$
lead to injected currents with different magnitudes but also with
vastly dissimilar angular dependences:\begin{multline}
\dot{\vec{J}}=2\Im\left[\eta_{I}^{xxxx}\right]E_{\omega}^{2}E_{2\omega}\sin(\Delta\phase)\\
\times\left[f(\theta,\yogh)\,\hat{\vec{e}}_{2\omega}+g(\theta,\yogh)\,\hat{\vec{e}}_{2\omega}^{\perp}\right],\label{eq:Jdot-linear-polarization-in-isotropic-medium}\end{multline}
where $f(\theta,\yogh)=\cos^{2}\theta+\yogh\,\sin^{2}\theta$ and
$g(\theta,\yogh)=\frac{1}{2}\left(1-\yogh\right)\,\sin2\theta$. Thus,
the current component that is parallel to $\hat{\vec{e}}_{2\omega}$
has a nonseparable dependence on $\theta$ and $\yogh$, whereas the
perpendicular component always follows $\sin2\theta$. Polar plots
of the functions $f(\theta,\yogh)$ and $g(\theta,\yogh)$ are shown
on the right-hand side of \prettyref{fig:yogh-bilayer} for $\yogh=-2$,
$-1$, $-0.5$, $0$, and $0.5$. Our result for graphene ($\yogh=-1$)
yields a clover-shaped angular distribution: the $\cos2\theta$ dependence
in \prettyref{eq:Jdot-linear-polarization-in-graphene}. For more
(or less) negative values of $\yogh$, the lobes around $\theta=\frac{\pi}{2}$
and $\frac{3\pi}{2}$ become more (or less) important. At $\yogh=0$,
these two lobes vanish; any current injected with perpendicular $\omega$
and $2\omega$ polarization axes is completely perpendicular to $\hat{\vec{e}}_{2\omega}$.
For $\yogh>0$ there are no nodes in the angular distribution. For
$\left|\yogh\right|>1$ the current parallel to $\hat{\vec{e}}_{2\omega}$
is stronger for perpendicular polarization axes compared to parallel
polarization axes. By scanning the photon energy in the range $\gamma_{1}<2\hbar\omega\lesssim3\gamma_{1}$,
the disparity parameter $d$ and thus the angular dependence of the
current injection in bilayer graphene vary significantly, in contrast
with the current injection in single-layer graphene. We note in particular
that the sharp changes in the value of $d$ near $2\hbar\omega\approx\gamma_{1}$
and $2\hbar\omega\approx2\gamma_{1}$ should be perceived experimentally
by rapid transitions in the angular dependence of the currents as
the photon energy is scanned.

\section{Summary and discussion\label{sec:Summary}}

We have calculated the response tensors for one- and two-photon carrier
injection and two-color current injection in graphene and bilayer
graphene. We find a strong, frequency-independent linear-circular
dichroism $\delta=-1$ in the two-photon response of graphene; for
the bilayer, $\delta$ also equals $-1$ when $2\hbar\omega<\gamma_{1}$
or $2\hbar\omega\gtrsim3\gamma_{1}$, and changes sign when $\gamma_{1}<2\hbar\omega\lesssim3\gamma_{1}$.
Using the optical CC technique, in-plane currents are generated for
co-circularly polarized and linearly-polarized beams. Such currents
could be detected in experiments making use of contacts, or by detecting
the emitted THz from the accelerated charges as was demonstrated by
Sun \emph{et~al.}\ in epitaxially-grown multilayer graphene \citep{Sun2010}.
In the bilayer, the dependence on the angle $\theta$ between linearly-polarized
light components at $\omega$ and $2\omega$ is strongly sensitive
to the photon energy for $\gamma_{1}<2\hbar\omega\lesssim3\gamma_{1}$.
This angular dependence is in sharp contrast to the prediction for
a single graphene layer, and could be mapped out experimentally as
a signature for interlayer coupling in epitaxially-grown multilayer
graphene samples, which are essentially thought of as uncoupled graphene
layers \citep{DeHeer2007}. We have assumed that the Fermi energy
is at the Dirac point, but for a nonzero Fermi energy $E_{F}$, our
predictions hold for $\hbar\omega>\left|E_{F}\right|$. When the Fermi
energy varies across an inhomogeneous sample, $E_{F}$ should be taken
as the largest local Fermi energy.

Our description of the bilayer considers only the strongest of the
interlayer coupling parameters, $\gamma_{1}$, and excludes the next-to-nearest
coupling parameters $\gamma_{3}$ and $\gamma_{4}$. These terms break
the isotropy of the model and introduce trigonal warping \citep{CastroNeto2009}.
However, we feel that excluding them in a first calculation is justified.
The inclusion of $\gamma_{3}$ hardly changes the conductivity spectrum
\citep{Min2009}. The same can be said of including the next-to-nearest
neighbor hoping term in the Hamiltonian for graphene \citep{Stauber2008}.

A natural extension of the current model is to consider AB-stacked
multilayer graphene samples. Koshino and Ando have shown that for
$n$ layers with $n$ even, the Hamiltonian can be decoupled into
$n/2$ bilayers (with $n$ odd, $(n-1)/2$ bilayers and one decoupled
single layer) \citep{Koshino2008a}. Each bilayer pair has a reduced
coupling strength $\lambda_{m}\gamma_{1}$ where $\lambda_{m}=2\cos\kappa_{m}$
and $\kappa_{m}$ is a wavevector in the stacking direction. The response
of the multilayer system is the sum of the bilayer systems responses
with detuned coupling strengths. In the limit of a high number of
layers, the wavevector $\kappa$ becomes a continuous variable and
there is a continuous spectrum of resonances, smearing out the response
tensor and eventually modeling the response of bulk graphite.

In the present paper, spin and valley degrees of freedom have contributed
only degeneracy factors of 2 to the injection tensors. But spin- or
valley-polarized currents could also be photogenerated if we first
lifted the spin or valley degeneracy, for example in a sample subjected
to a magnetic or pseudomagnetic field, provided that the Fermi energy
is chosen appropriately. Strain-induced pseudomagnetic fields have
been suggested to lift the degeneracy of the K and K' points by 100\,meV
\citep{Guinea2010a,Guinea2010b}.

We conclude by pointing out the differences between coherent current
control in graphene and in conventional semiconductors such as GaAs.
In two-color CC experiments in gapped semiconductors, the fields are
typically chosen so that the semiconducting bandgap $E_{g}$ lies
between $\hbar\omega<E_{g}<2\hbar\omega$. Thus, one-photon absorption
at the fundamental frequency is energetically forbidden. However,
since the band dispersions of graphene and bilayer graphene are gapless,
for a clean, unbiased sample there is a nonzero joint density of states
down to zero frequency and one-photon absorption is always present.
This raises an issue for two-color CC experiments where the usual
best practice is to have balanced absorption between the first-order
process at $2\omega$ and the second-order process at $\omega$. To
achieve this, the fundamental beam is given most of the power. If
this beam is absorbed in the linear regime, it can potentially flood
the sample with carriers that are not taking part in the quantum interference.
We note however that this has not led to difficulties in observing
the coherent current control in multilayer epitaxial graphene \citep{Sun2010}.

The opening of a sufficiently large gap in the band dispersion of
graphene would completely eliminate one-photon absorption at $\omega$.
The weak spin-orbit coupling offers only a very small gap that has
been calculated to be on the $\upmu$eV scale for graphene \citep{Min2006,Gmitra2009}
and bilayer graphene \citep{Guinea2010c}, while gaps induced by the
substrate \citep{McCann2006b,Nicol2008,Enderlein2010} or by confinement
\citep{Brey2006b,Han2007} (in certain nanoribbon geometries, similar
to the way carbon nanotubes can acquire a gap) are typically tens
of meV. Most interestingly, in bilayer graphene gap opening can also
occur due to $z$-axis asymmetry between the two layers, which can
be field-induced \citep{Guo2008}. Field-induced gaps are tunable
up to hundreds of meV.

The problem of linear absorption of the fundamental could also be
circumvented by a nonzero Fermi energy $E_{F}$, taking advantage
of Pauli blocking to prevent one-photon absorption at $\omega$. With
the relation $\hbar\omega/2<\left|E_{F}\right|<\hbar\omega$, one
effectively has the same condition as typical gapped semiconductors.
This offers even more tunability since the Fermi energy can be gate-controlled,
and could lead to novel electro-optical devices making use of coherent
current control.

\appendix

\begin{acknowledgments}
This work was supported by FQRNT, by DFG under project numbers SFB
767 and FOR 912, and by CAP and NSERC. The authors acknowledge useful
discussions with Dong Sun and Ted Norris.
\end{acknowledgments}

%


\begin{thebibliography}{36}%
\makeatletter
\providecommand \@ifxundefined [1]{%
 \@ifx{#1\undefined}
}%
\providecommand \@ifnum [1]{%
 \ifnum #1\expandafter \@firstoftwo
 \else \expandafter \@secondoftwo
 \fi
}%
\providecommand \@ifx [1]{%
 \ifx #1\expandafter \@firstoftwo
 \else \expandafter \@secondoftwo
 \fi
}%
\providecommand \natexlab [1]{#1}%
\providecommand \bibnamefont  [1]{#1}%
\providecommand \bibfnamefont [1]{#1}%
\providecommand \citenamefont [1]{#1}%
\providecommand \href@noop [0]{\@secondoftwo}%
\providecommand \href [0]{\begingroup \@sanitize@url \@href}%
\providecommand \@href[1]{\@@startlink{#1}\@@href}%
\providecommand \@@href[1]{\endgroup#1\@@endlink}%
\providecommand \@sanitize@url [0]{\catcode `\\12\catcode `\$12\catcode
  `\&12\catcode `\#12\catcode `\^12\catcode `\_12\catcode `\%12\relax}%
\providecommand \@@startlink[1]{}%
\providecommand \@@endlink[0]{}%
\long\def\true@sw#1#2{#1}%
\long\def\false@sw#1#2{#2}%
\@ifxundefined\pdfoutput%
{\true@sw}{\@ifnum{\z@=\pdfoutput}{\true@sw}{\false@sw}}%
{
\def\@@startlink#1{\leavevmode%
\color\@urlcolor}%
\def\@@endlink{\color{black}}%
}{
\gdef\@@startlink#1{%
  \leavevmode
  \pdfstartlink\pdfstartlink@attr
   user{/Subtype/Link/A<</Type/Action/S/URI/URI(#1)>>}%
  \relax
  \color\@urlcolor
}%
\gdef\pdfstartlink@attr{%
      attr{%
        \Hy@setpdfborder
        \ifx\@pdfhightlight\@empty
        \else
          /H\@pdfhighlight
        \fi
        \ifx\@urlbordercolor\relax
        \else
          /C[\@urlbordercolor]%
        \fi
      }%
}
\gdef\@@endlink{%
\color{black}%
\pdfendlink%
}%
}%
\def\firstpage#1--#2\relax{#1}
\def\pages#1{\firstpage #1\relax}
\providecommand \url  [0]{\begingroup\@sanitize@url \@url }%
\providecommand \@url [1]{\endgroup\@href {#1}{\urlprefix }}%
\providecommand \urlprefix  [0]{URL }%
\providecommand \Eprint [0]{\href }%
\@ifxundefined \urlstyle {%
  \providecommand \doi  [0]{\begingroup \@sanitize@url \@doi}%
  \providecommand \@doi [1]{\endgroup \@@startlink {\doibase
  #1}doi:\discretionary {}{}{}#1\@@endlink }%
}{%
  \providecommand \doi  [0]{doi:\discretionary{}{}{}\begingroup
  \urlstyle{rm}\Url }%
}%
\providecommand \doibase [0]{http://dx.doi.org/}%
\providecommand \Doi [0]{\begingroup \@sanitize@url \@Doi }%
\providecommand \@Doi  [1]{\endgroup\@@startlink{\doibase#1}\@@Doi}%
\providecommand \@@Doi [1]{#1\@@endlink}%
\providecommand \selectlanguage [0]{\@gobble}%
\renewcommand \@bibibid@[1]{\emph{ibid}.,}%
\renewcommand \@bib@Xauthor[1]{%
  \let\@bib@Xjournal\@gobble%
  \let\@bib@Xtitle\@gobble%
}%
\newcommand \@bib@Yauthor{%
  \def\@bib@Xjournal##1{%
    \begingroup%
      \let\bibinfo@X@journal%
      \@bib@Z@journal##1%
    \endgroup}%
}%
\providecommand \bibinfo  [0]{\@secondoftwo}%
\providecommand \bibfield  [0]{\@secondoftwo}%
\providecommand \translation [1]{[#1]}%
\providecommand \BibitemOpen [0]{}%
\providecommand \bibitemStop [0]{}%
\providecommand \bibitemNoStop [0]{.\EOS\space}%
\providecommand \EOS [0]{\spacefactor3000\relax}%
\providecommand \BibitemShut  [1]{\csname bibitem#1\endcsname}%
\bibitem [{\citenamefont {Novoselov} \emph {et~al.}(2004)\citenamefont
  {Novoselov}, \citenamefont {Geim}, \citenamefont {Morozov}, \citenamefont
  {Jiang}, \citenamefont {Zhang}, \citenamefont {Dubonos}, \citenamefont
  {Grigorieva}, and \citenamefont {Firsov}}]{Novoselov2004}
  \BibitemOpen
  \bibfield  {author} {\bibinfo {author} {\bibfnamefont {K.~S.} \bibnamefont
  {Novoselov}}, \bibinfo {author} {\bibfnamefont {A.~K.} \bibnamefont {Geim}},
  \bibinfo {author} {\bibfnamefont {S.~V.} \bibnamefont {Morozov}}, \bibinfo
  {author} {\bibfnamefont {D.}~\bibnamefont {Jiang}}, \bibinfo {author}
  {\bibfnamefont {Y.}~\bibnamefont {Zhang}}, \bibinfo {author} {\bibfnamefont
  {S.~V.} \bibnamefont {Dubonos}}, \bibinfo {author} {\bibfnamefont {I.~V.}
  \bibnamefont {Grigorieva}},  and \bibinfo {author} {\bibfnamefont {A.~A.}
  \bibnamefont {Firsov}}, }\Doi {10.1126/science.1102896} {\bibfield  {journal}
  {\bibinfo  {journal} {Science}\ }\textbf {\bibinfo {volume} {306}}, \bibinfo
  {pages} {666} (\bibinfo {year} {2004})}\bibfield  {note} {; preprint at
  \Eprint {http://arxiv.org/abs/arXiv:cond-mat/0410550} {\bibinfo {eprint}
  {arXiv:cond-mat/0410550}}}\BibitemShut {NoStop}%
\bibitem [{\citenamefont {Novoselov} \emph
  {et~al.}(2005){\natexlab{a}}\citenamefont {Novoselov}, \citenamefont {Jiang},
  \citenamefont {Schedin}, \citenamefont {Booth}, \citenamefont {Khotkevich},
  \citenamefont {Morozov}, and \citenamefont {Geim}}]{Novoselov2005a}
  \BibitemOpen
  \bibfield  {author} {\bibinfo {author} {\bibfnamefont {K.~S.} \bibnamefont
  {Novoselov}}, \bibinfo {author} {\bibfnamefont {D.}~\bibnamefont {Jiang}},
  \bibinfo {author} {\bibfnamefont {F.}~\bibnamefont {Schedin}}, \bibinfo
  {author} {\bibfnamefont {T.~J.} \bibnamefont {Booth}}, \bibinfo {author}
  {\bibfnamefont {V.~V.} \bibnamefont {Khotkevich}}, \bibinfo {author}
  {\bibfnamefont {S.~V.} \bibnamefont {Morozov}},  and \bibinfo {author}
  {\bibfnamefont {A.~K.} \bibnamefont {Geim}}, }\Doi {10.1073/pnas.0502848102}
  {\bibfield  {journal} {\bibinfo  {journal} {Proc. Natl. Acad. Sci. USA}\
  }\textbf {\bibinfo {volume} {102}}, \bibinfo {pages} {10451} (\bibinfo {year}
  {2005}{\natexlab{a}})}\bibfield  {note} {; preprint at \Eprint
  {http://arxiv.org/abs/arXiv:cond-mat/0503533} {\bibinfo {eprint}
  {arXiv:cond-mat/0503533}}}\BibitemShut {NoStop}%
\bibitem [{\citenamefont {Novoselov} \emph
  {et~al.}(2005){\natexlab{b}}\citenamefont {Novoselov}, \citenamefont {Geim},
  \citenamefont {Morozov}, \citenamefont {Jiang}, \citenamefont {Katsnelson},
  \citenamefont {Grigorieva}, \citenamefont {Dubonos}, and \citenamefont
  {Firsov}}]{Novoselov2005b}
  \BibitemOpen
  \bibfield  {author} {\bibinfo {author} {\bibfnamefont {K.~S.} \bibnamefont
  {Novoselov}}, \bibinfo {author} {\bibfnamefont {A.~K.} \bibnamefont {Geim}},
  \bibinfo {author} {\bibfnamefont {S.~V.} \bibnamefont {Morozov}}, \bibinfo
  {author} {\bibfnamefont {D.}~\bibnamefont {Jiang}}, \bibinfo {author}
  {\bibfnamefont {M.~I.} \bibnamefont {Katsnelson}}, \bibinfo {author}
  {\bibfnamefont {I.~V.} \bibnamefont {Grigorieva}}, \bibinfo {author}
  {\bibfnamefont {S.~V.} \bibnamefont {Dubonos}},  and \bibinfo {author}
  {\bibfnamefont {A.~A.} \bibnamefont {Firsov}}, }\Doi {10.1038/nature04233}
  {\bibfield  {journal} {\bibinfo  {journal} {Nature}\ }\textbf {\bibinfo
  {volume} {438}}, \bibinfo {pages} {197} (\bibinfo {year}
  {2005}{\natexlab{b}})}\bibfield  {note} {; preprint at \Eprint
  {http://arxiv.org/abs/arXiv:cond-mat/0509330} {\bibinfo {eprint}
  {arXiv:cond-mat/0509330}}}\BibitemShut {NoStop}%
\bibitem [{\citenamefont {Geim} and \citenamefont {Novoselov}(2007)}]{Geim2007}
  \BibitemOpen
  \bibfield  {author} {\bibinfo {author} {\bibfnamefont {A.~K.} \bibnamefont
  {Geim}} and \bibinfo {author} {\bibfnamefont {K.~S.} \bibnamefont
  {Novoselov}}, }\Doi {10.1038/nmat1849} {\bibfield  {journal} {\bibinfo
  {journal} {Nat. Mater.}\ }\textbf {\bibinfo {volume} {6}}, \bibinfo {pages}
  {183} (\bibinfo {year} {2007})}\bibfield  {note} {; preprint at \Eprint
  {http://arxiv.org/abs/arXiv:cond-mat/0702595} {\bibinfo {eprint}
  {arXiv:cond-mat/0702595}}}\BibitemShut {NoStop}%
\bibitem [{\citenamefont {Castro~Neto} \emph {et~al.}(2009)\citenamefont
  {Castro~Neto}, \citenamefont {Guinea}, \citenamefont {Peres}, \citenamefont
  {Novoselov}, and \citenamefont {Geim}}]{CastroNeto2009}
  \BibitemOpen
  \bibfield  {author} {\bibinfo {author} {\bibfnamefont {A.~H.} \bibnamefont
  {Castro~Neto}}, \bibinfo {author} {\bibfnamefont {F.}~\bibnamefont {Guinea}},
  \bibinfo {author} {\bibfnamefont {N.~M.~R.} \bibnamefont {Peres}}, \bibinfo
  {author} {\bibfnamefont {K.~S.} \bibnamefont {Novoselov}},  and \bibinfo
  {author} {\bibfnamefont {A.~K.} \bibnamefont {Geim}}, }\Doi
  {10.1103/RevModPhys.81.109} {\bibfield  {journal} {\bibinfo  {journal} {Rev.
  Mod. Phys.}\ }\textbf {\bibinfo {volume} {81}}, \bibinfo {eid} {109}
  (\bibinfo {year} {2009})}\bibfield  {note} {; preprint at \Eprint
  {http://arxiv.org/abs/arXiv:0709.1163} {\bibinfo {eprint}
  {arXiv:0709.1163}}}\BibitemShut {NoStop}%
\bibitem [{\citenamefont {Nair} \emph {et~al.}(2008)\citenamefont {Nair},
  \citenamefont {Blake}, \citenamefont {Grigorenko}, \citenamefont {Novoselov},
  \citenamefont {Booth}, \citenamefont {Stauber}, \citenamefont {Peres}, and
  \citenamefont {Geim}}]{Nair2008}
  \BibitemOpen
  \bibfield  {author} {\bibinfo {author} {\bibfnamefont {R.~R.} \bibnamefont
  {Nair}}, \bibinfo {author} {\bibfnamefont {P.}~\bibnamefont {Blake}},
  \bibinfo {author} {\bibfnamefont {A.~N.} \bibnamefont {Grigorenko}}, \bibinfo
  {author} {\bibfnamefont {K.~S.} \bibnamefont {Novoselov}}, \bibinfo {author}
  {\bibfnamefont {T.~J.} \bibnamefont {Booth}}, \bibinfo {author}
  {\bibfnamefont {T.}~\bibnamefont {Stauber}}, \bibinfo {author} {\bibfnamefont
  {N.~M.~R.} \bibnamefont {Peres}},  and \bibinfo {author} {\bibfnamefont
  {A.~K.} \bibnamefont {Geim}}, }\Doi {10.1126/science.1156965} {\bibfield
  {journal} {\bibinfo  {journal} {Science}\ }\textbf {\bibinfo {volume} {320}},
  \bibinfo {pages} {1308} (\bibinfo {year} {2008})}\bibfield  {note} {;
  preprint at \Eprint {http://arxiv.org/abs/arXiv:0803.3718} {\bibinfo {eprint}
  {arXiv:0803.3718}}}\BibitemShut {NoStop}%
\bibitem [{\citenamefont {Novoselov} \emph {et~al.}(2006)\citenamefont
  {Novoselov}, \citenamefont {McCann}, \citenamefont {Morozov}, \citenamefont
  {Fal'ko}, \citenamefont {Katsnelson}, \citenamefont {Zeitler}, \citenamefont
  {Jiang}, \citenamefont {Schedin}, and \citenamefont {Geim}}]{Novoselov2006}
  \BibitemOpen
  \bibfield  {author} {\bibinfo {author} {\bibfnamefont {K.~S.} \bibnamefont
  {Novoselov}}, \bibinfo {author} {\bibfnamefont {E.}~\bibnamefont {McCann}},
  \bibinfo {author} {\bibfnamefont {S.~V.} \bibnamefont {Morozov}}, \bibinfo
  {author} {\bibfnamefont {V.~I.} \bibnamefont {Fal'ko}}, \bibinfo {author}
  {\bibfnamefont {M.~I.} \bibnamefont {Katsnelson}}, \bibinfo {author}
  {\bibfnamefont {U.}~\bibnamefont {Zeitler}}, \bibinfo {author} {\bibfnamefont
  {D.}~\bibnamefont {Jiang}}, \bibinfo {author} {\bibfnamefont
  {F.}~\bibnamefont {Schedin}},  and \bibinfo {author} {\bibfnamefont {A.~K.}
  \bibnamefont {Geim}}, }\Doi {10.1038/nphys245} {\bibfield  {journal}
  {\bibinfo  {journal} {Nat. Phys.}\ }\textbf {\bibinfo {volume} {2}}, \bibinfo
  {pages} {177} (\bibinfo {year} {2006})}\bibfield  {note} {; preprint at
  \Eprint {http://arxiv.org/abs/arXiv:cond-mat/0602565} {\bibinfo {eprint}
  {arXiv:cond-mat/0602565}}}\BibitemShut {NoStop}%
\bibitem [{\citenamefont {Katsnelson}(2006)}]{Katsnelson2006a}
  \BibitemOpen
  \bibfield  {author} {\bibinfo {author} {\bibfnamefont {M.~I.} \bibnamefont
  {Katsnelson}}, }\Doi {10.1140/epjb/e2006-00294-6} {\bibfield  {journal}
  {\bibinfo  {journal} {Eur. Phys. J. B}\ }\textbf {\bibinfo {volume} {52}},
  \bibinfo {pages} {151} (\bibinfo {year} {2006})}\bibfield  {note} {; preprint
  at \Eprint {http://arxiv.org/abs/arXiv:cond-mat/0606611} {\bibinfo {eprint}
  {arXiv:cond-mat/0606611}}}\BibitemShut {NoStop}%
\bibitem [{\citenamefont {Morozov} \emph {et~al.}(2008)\citenamefont {Morozov},
  \citenamefont {Novoselov}, \citenamefont {Katsnelson}, \citenamefont
  {Schedin}, \citenamefont {Elias}, \citenamefont {Jaszczak}, and \citenamefont
  {Geim}}]{Morozov2008a}
  \BibitemOpen
  \bibfield  {author} {\bibinfo {author} {\bibfnamefont {S.~V.} \bibnamefont
  {Morozov}}, \bibinfo {author} {\bibfnamefont {K.~S.} \bibnamefont
  {Novoselov}}, \bibinfo {author} {\bibfnamefont {M.~I.} \bibnamefont
  {Katsnelson}}, \bibinfo {author} {\bibfnamefont {F.}~\bibnamefont {Schedin}},
  \bibinfo {author} {\bibfnamefont {D.~C.} \bibnamefont {Elias}}, \bibinfo
  {author} {\bibfnamefont {J.~A.} \bibnamefont {Jaszczak}},  and \bibinfo
  {author} {\bibfnamefont {A.~K.} \bibnamefont {Geim}}, }\Doi
  {10.1103/PhysRevLett.100.016602} {\bibfield  {journal} {\bibinfo  {journal}
  {Phys. Rev. Lett.}\ }\textbf {\bibinfo {volume} {100}}, \bibinfo {pages}
  {016602} (\bibinfo {year} {2008})}\bibfield  {note} {; preprint at \Eprint
  {http://arxiv.org/abs/arXiv:0710.5304} {\bibinfo {eprint}
  {arXiv:0710.5304}}}\BibitemShut {NoStop}%
\bibitem [{\citenamefont {Rabitz}(2005)}]{coherent-control-review}
  \BibitemOpen
  \bibfield  {author} {\bibinfo {author} {\bibfnamefont {H.}~\bibnamefont
  {Rabitz}}, }in \Doi {10.1016/B0-12-369395-0/00903-9} {\emph {\bibinfo
  {booktitle} {Encyclopedia of Modern Optics}}}\  (\bibinfo  {publisher}
  {Elsevier}, \bibinfo {address} {Oxford}, \bibinfo {year} {2005}) pp. \bibinfo
  {pages} {123--133};
\newblock \bibinfo {note} {{R}.~J. Lewis, \emph{ibid.}, pp.~133--137; H.~M. van
  Driel and J.~E. Sipe, \emph{ibid.}, pp.~137--143}\BibitemShut {NoStop}%
\bibitem [{\citenamefont {Atanasov} \emph {et~al.}(1996)\citenamefont
  {Atanasov}, \citenamefont {Hach{\'e}}, \citenamefont {Hughes}, \citenamefont
  {van Driel}, and \citenamefont {Sipe}}]{Atanasov1996}
  \BibitemOpen
  \bibfield  {author} {\bibinfo {author} {\bibfnamefont {R.}~\bibnamefont
  {Atanasov}}, \bibinfo {author} {\bibfnamefont {A.}~\bibnamefont {Hach{\'e}}},
  \bibinfo {author} {\bibfnamefont {J.~L.~P.} \bibnamefont {Hughes}}, \bibinfo
  {author} {\bibfnamefont {H.~M.} \bibnamefont {van Driel}},  and \bibinfo
  {author} {\bibfnamefont {J.~E.} \bibnamefont {Sipe}}, }\Doi
  {10.1103/PhysRevLett.76.1703} {\bibfield  {journal} {\bibinfo  {journal}
  {Phys. Rev. Lett.}\ }\textbf {\bibinfo {volume} {76}}, \bibinfo {pages}
  {1703} (\bibinfo {year} {1996})}\BibitemShut {NoStop}%
\bibitem [{\citenamefont {Mele} \emph {et~al.}(2000)\citenamefont {Mele},
  \citenamefont {Kr{\'a}l}, and \citenamefont {Tom{\'a}nek}}]{Mele2000}
  \BibitemOpen
  \bibfield  {author} {\bibinfo {author} {\bibfnamefont {E.~J.} \bibnamefont
  {Mele}}, \bibinfo {author} {\bibfnamefont {P.}~\bibnamefont {Kr{\'a}l}},  and
  \bibinfo {author} {\bibfnamefont {D.}~\bibnamefont {Tom{\'a}nek}}, }\Doi
  {10.1103/PhysRevB.61.7669} {\bibfield  {journal} {\bibinfo  {journal} {Phys.
  Rev. B}\ }\textbf {\bibinfo {volume} {61}}, \bibinfo {pages} {7669} (\bibinfo
  {year} {2000})}\bibfield  {note} {; preprint at \Eprint
  {http://arxiv.org/abs/arXiv:cond-mat/9911151} {\bibinfo {eprint}
  {arXiv:cond-mat/9911151}}}\BibitemShut {NoStop}%
\bibitem [{\citenamefont {Sun} \emph {et~al.}(2010)\citenamefont {Sun},
  \citenamefont {Divin}, \citenamefont {Rioux}, \citenamefont {Sipe},
  \citenamefont {Berger}, \citenamefont {de~Heer}, \citenamefont {First}, and
  \citenamefont {Norris}}]{Sun2010}
  \BibitemOpen
  \bibfield  {author} {\bibinfo {author} {\bibfnamefont {D.}~\bibnamefont
  {Sun}}, \bibinfo {author} {\bibfnamefont {C.}~\bibnamefont {Divin}}, \bibinfo
  {author} {\bibfnamefont {J.}~\bibnamefont {Rioux}}, \bibinfo {author}
  {\bibfnamefont {J.~E.} \bibnamefont {Sipe}}, \bibinfo {author} {\bibfnamefont
  {C.}~\bibnamefont {Berger}}, \bibinfo {author} {\bibfnamefont {W.~A.}
  \bibnamefont {de~Heer}}, \bibinfo {author} {\bibfnamefont {P.~N.}
  \bibnamefont {First}},  and \bibinfo {author} {\bibfnamefont {{\relax
  Th}.~B.} \bibnamefont {Norris}}, }\Doi {10.1021/nl9040737} {\bibfield
  {journal} {\bibinfo  {journal} {Nano Lett.}\ }\textbf {\bibinfo {volume}
  {10}}, \bibinfo {pages} {1293} (\bibinfo {year} {2010})}\BibitemShut
  {NoStop}%
\bibitem [{\citenamefont {Newson} \emph {et~al.}(2008)\citenamefont {Newson},
  \citenamefont {M{\'e}nard}, \citenamefont {Sames}, \citenamefont {Betz}, and
  \citenamefont {van Driel}}]{Newson2008}
  \BibitemOpen
  \bibfield  {author} {\bibinfo {author} {\bibfnamefont {R.~W.} \bibnamefont
  {Newson}}, \bibinfo {author} {\bibfnamefont {J.-M.} \bibnamefont
  {M{\'e}nard}}, \bibinfo {author} {\bibfnamefont {C.}~\bibnamefont {Sames}},
  \bibinfo {author} {\bibfnamefont {M.}~\bibnamefont {Betz}},  and \bibinfo
  {author} {\bibfnamefont {H.~M.} \bibnamefont {van Driel}}, }\Doi
  {10.1021/nl073305l} {\bibfield  {journal} {\bibinfo  {journal} {Nano Lett.}\
  }\textbf {\bibinfo {volume} {8}}, \bibinfo {pages} {1586} (\bibinfo {year}
  {2008})}\BibitemShut {NoStop}%
\bibitem [{\citenamefont {Nicol} and \citenamefont
  {Carbotte}(2008)}]{Nicol2008}
  \BibitemOpen
  \bibfield  {author} {\bibinfo {author} {\bibfnamefont {E.~J.} \bibnamefont
  {Nicol}} and \bibinfo {author} {\bibfnamefont {J.~P.} \bibnamefont
  {Carbotte}}, }\Doi {10.1103/PhysRevB.77.155409} {\bibfield  {journal}
  {\bibinfo  {journal} {Phys. Rev. B}\ }\textbf {\bibinfo {volume} {77}},
  \bibinfo {eid} {155409} (\bibinfo {year} {2008})}\bibfield  {note} {;
  preprint at \Eprint {http://arxiv.org/abs/arXiv:0801.1836} {\bibinfo {eprint}
  {arXiv:0801.1836}}}\BibitemShut {NoStop}%
\bibitem [{\citenamefont {van Driel} and \citenamefont
  {Sipe}(2000)}]{VanDriel2000}
  \BibitemOpen
  \bibfield  {author} {\bibinfo {author} {\bibfnamefont {H.~M.} \bibnamefont
  {van Driel}} and \bibinfo {author} {\bibfnamefont {J.~E.} \bibnamefont
  {Sipe}}, }in \href@noop {} {\emph {\bibinfo {booktitle} {Ultrafast Phenomena
  in Semiconductors}}}, \bibinfo {editor} {edited by \bibinfo {editor}
  {\bibfnamefont {K.-T.} \bibnamefont {Tsen}}}\  (\bibinfo  {publisher}
  {Springer-Verlag}, \bibinfo {address} {Berlin}, \bibinfo {year} {2000})
  Chap.~\bibinfo {chapter} {5}, pp. \bibinfo {pages} {261--306}\BibitemShut
  {NoStop}%
\bibitem [{\citenamefont {Hutchings} and \citenamefont
  {Wherrett}(1994)}]{note-on-isotropic-model}
  \BibitemOpen
  \bibfield  {note} {\bibinfo {note} {Other nonzero components of the isotropic
  model are $\xi^{xyxy}=\xi^{xyyx}=\frac{1}{2}\xi^{xxxx}(1-\delta)$, and those
  obtained by $x\leftrightarrow y$ permutations, omitted throughout for
  simplicity; see~}}\bibfield  {author} {\bibinfo {author} {\bibfnamefont
  {D.~C.} \bibnamefont {Hutchings}} and \bibinfo {author} {\bibfnamefont
  {B.~S.} \bibnamefont {Wherrett}}, }\Doi {10.1016/0925-3467(94)90029-9}
  {\bibfield  {journal} {\bibinfo  {journal} {Opt. Mater.}\ }\textbf {\bibinfo
  {volume} {3}}, \bibinfo {pages} {53} (\bibinfo {year} {1994})}\BibitemShut
  {NoStop}%
\bibitem [{\citenamefont {Gusynin} \emph {et~al.}(2006)\citenamefont {Gusynin},
  \citenamefont {Sharapov}, and \citenamefont {Carbotte}}]{Gusynin2006b}
  \BibitemOpen
  \bibfield  {author} {\bibinfo {author} {\bibfnamefont {V.~P.} \bibnamefont
  {Gusynin}}, \bibinfo {author} {\bibfnamefont {S.~G.} \bibnamefont
  {Sharapov}},  and \bibinfo {author} {\bibfnamefont {J.~P.} \bibnamefont
  {Carbotte}}, }\Doi {10.1103/PhysRevLett.96.256802} {\bibfield  {journal}
  {\bibinfo  {journal} {Phys. Rev. Lett.}\ }\textbf {\bibinfo {volume} {96}},
  \bibinfo {pages} {256802} (\bibinfo {year} {2006})}\bibfield  {note} {;
  preprint at \Eprint {http://arxiv.org/abs/arXiv:cond-mat/0603267} {\bibinfo
  {eprint} {arXiv:cond-mat/0603267}}}\BibitemShut {NoStop}%
\bibitem [{Note1()}]{Note1}
  \BibitemOpen
  \bibinfo {note} {Looking at how the interference term in the carrier
  injection transforms under the action of inversion symmetry, one gets
  $\protect \mathaccentV {dot}05F{n}_{I}=\xi _{I}^{abc}(\omega )E^{a*}(\omega
  )E^{b*}(\omega )E^{c}(2\omega )+\protect \mathrm {c.c.}\rightarrow \protect
  \mathaccentV {dot}05F{n}_{I}=-\xi _{I}^{abc}(\omega )E^{a*}(\omega
  )E^{b*}(\omega )E^{c}(2\omega )+\protect \mathrm {c.c.}$, which imposes $\xi
  _{I}=0$.}\BibitemShut {Stop}%
\bibitem [{\citenamefont {Najmaie} \emph {et~al.}(2003)\citenamefont {Najmaie},
  \citenamefont {Bhat}, and \citenamefont {Sipe}}]{Najmaie2003}
  \BibitemOpen
  \bibfield  {author} {\bibinfo {author} {\bibfnamefont {A.}~\bibnamefont
  {Najmaie}}, \bibinfo {author} {\bibfnamefont {R.~D.~R.} \bibnamefont {Bhat}},
   and \bibinfo {author} {\bibfnamefont {J.~E.} \bibnamefont {Sipe}}, }\Doi
  {10.1103/PhysRevB.68.165348} {\bibfield  {journal} {\bibinfo  {journal}
  {Phys. Rev. B}\ }\textbf {\bibinfo {volume} {68}}, \bibinfo {pages} {165348}
  (\bibinfo {year} {2003})}\BibitemShut {NoStop}%
\bibitem [{\citenamefont {Bhat} and \citenamefont {Sipe}(2006)}]{Bhat2006}
  \BibitemOpen
  \bibfield  {author} {\bibinfo {author} {\bibfnamefont {R.~D.~R.} \bibnamefont
  {Bhat}} and \bibinfo {author} {\bibfnamefont {J.~E.} \bibnamefont {Sipe}},
  }\bibinfo {note} {\href {http://arxiv.org/abs/arXiv:cond-mat/0601277}
  {arXiv:cond-mat/0601277}}\ (unpublished)\BibitemShut {NoStop}%
\bibitem [{\citenamefont {Abergel} and \citenamefont
  {Fal'ko}(2007)}]{Abergel2007}
  \BibitemOpen
  \bibfield  {author} {\bibinfo {author} {\bibfnamefont {D.~S.~L.} \bibnamefont
  {Abergel}} and \bibinfo {author} {\bibfnamefont {V.~I.} \bibnamefont
  {Fal'ko}}, }\Doi {10.1103/PhysRevB.75.155430} {\bibfield  {journal} {\bibinfo
   {journal} {Phys. Rev. B}\ }\textbf {\bibinfo {volume} {75}}, \bibinfo
  {pages} {155430} (\bibinfo {year} {2007})}\bibfield  {note} {; preprint at
  \Eprint {http://arxiv.org/abs/arXiv:cond-mat/0610673} {\bibinfo {eprint}
  {arXiv:cond-mat/0610673}}}\BibitemShut {NoStop}%
\bibitem [{\citenamefont {de~Heer} \emph {et~al.}(2007)\citenamefont {de~Heer},
  \citenamefont {Berger}, \citenamefont {Wu}, \citenamefont {First},
  \citenamefont {Conrad}, \citenamefont {Li}, \citenamefont {Li}, \citenamefont
  {Sprinkle}, \citenamefont {Hass}, \citenamefont {Sadowski}, \citenamefont
  {Potemski}, and \citenamefont {Martinez}}]{DeHeer2007}
  \BibitemOpen
  \bibfield  {author} {\bibinfo {author} {\bibfnamefont {W.~A.} \bibnamefont
  {de~Heer}}, \bibinfo {author} {\bibfnamefont {C.}~\bibnamefont {Berger}},
  \bibinfo {author} {\bibfnamefont {X.}~\bibnamefont {Wu}}, \bibinfo {author}
  {\bibfnamefont {P.~N.} \bibnamefont {First}}, \bibinfo {author}
  {\bibfnamefont {E.~H.} \bibnamefont {Conrad}}, \bibinfo {author}
  {\bibfnamefont {X.}~\bibnamefont {Li}}, \bibinfo {author} {\bibfnamefont
  {T.}~\bibnamefont {Li}}, \bibinfo {author} {\bibfnamefont {M.}~\bibnamefont
  {Sprinkle}}, \bibinfo {author} {\bibfnamefont {J.}~\bibnamefont {Hass}},
  \bibinfo {author} {\bibfnamefont {M.~L.} \bibnamefont {Sadowski}}, \bibinfo
  {author} {\bibfnamefont {M.}~\bibnamefont {Potemski}},  and \bibinfo {author}
  {\bibfnamefont {G.}~\bibnamefont {Martinez}}, }\Doi
  {10.1016/j.ssc.2007.04.023} {\bibfield  {journal} {\bibinfo  {journal} {Solid
  State Commun.}\ }\textbf {\bibinfo {volume} {143}}, \bibinfo {pages} {92}
  (\bibinfo {year} {2007})}\bibfield  {note} {; preprint at \Eprint
  {http://arxiv.org/abs/arXiv:0704.0285} {\bibinfo {eprint}
  {arXiv:0704.0285}}}\BibitemShut {NoStop}%
\bibitem [{\citenamefont {Min} and \citenamefont {MacDonald}(2009)}]{Min2009}
  \BibitemOpen
  \bibfield  {author} {\bibinfo {author} {\bibfnamefont {H.}~\bibnamefont
  {Min}} and \bibinfo {author} {\bibfnamefont {A.~H.} \bibnamefont
  {MacDonald}}, }\Doi {10.1103/PhysRevLett.103.067402} {\bibfield  {journal}
  {\bibinfo  {journal} {Phys. Rev. Lett.}\ }\textbf {\bibinfo {volume} {103}},
  \bibinfo {eid} {067402} (\bibinfo {year} {2009})}\bibfield  {note} {;
  preprint at \Eprint {http://arxiv.org/abs/arXiv:0903.2163} {\bibinfo {eprint}
  {arXiv:0903.2163}}}\BibitemShut {NoStop}%
\bibitem [{\citenamefont {Stauber} \emph {et~al.}(2008)\citenamefont {Stauber},
  \citenamefont {Peres}, and \citenamefont {Geim}}]{Stauber2008}
  \BibitemOpen
  \bibfield  {author} {\bibinfo {author} {\bibfnamefont {T.}~\bibnamefont
  {Stauber}}, \bibinfo {author} {\bibfnamefont {N.~M.~R.} \bibnamefont
  {Peres}},  and \bibinfo {author} {\bibfnamefont {A.~K.} \bibnamefont {Geim}},
  }\Doi {10.1103/PhysRevB.78.085432} {\bibfield  {journal} {\bibinfo  {journal}
  {Phys. Rev. B}\ }\textbf {\bibinfo {volume} {78}}, \bibinfo {eid} {085432}
  (\bibinfo {year} {2008})}\bibfield  {note} {; preprint at \Eprint
  {http://arxiv.org/abs/arXiv:0803.1802} {\bibinfo {eprint}
  {arXiv:0803.1802}}}\BibitemShut {NoStop}%
\bibitem [{\citenamefont {Koshino} and \citenamefont
  {Ando}(2008)}]{Koshino2008a}
  \BibitemOpen
  \bibfield  {author} {\bibinfo {author} {\bibfnamefont {M.}~\bibnamefont
  {Koshino}} and \bibinfo {author} {\bibfnamefont {T.}~\bibnamefont {Ando}},
  }\Doi {10.1103/PhysRevB.77.115313} {\bibfield  {journal} {\bibinfo  {journal}
  {Phys. Rev. B}\ }\textbf {\bibinfo {volume} {77}}, \bibinfo {pages} {115313}
  (\bibinfo {year} {2008})}\bibfield  {note} {; preprint at \Eprint
  {http://arxiv.org/abs/arXiv:0803.3023} {\bibinfo {eprint}
  {arXiv:0803.3023}}}\BibitemShut {NoStop}%
\bibitem [{\citenamefont {Guinea} \emph
  {et~al.}(2010){\natexlab{a}}\citenamefont {Guinea}, \citenamefont
  {Katsnelson}, and \citenamefont {Geim}}]{Guinea2010a}
  \BibitemOpen
  \bibfield  {author} {\bibinfo {author} {\bibfnamefont {F.}~\bibnamefont
  {Guinea}}, \bibinfo {author} {\bibfnamefont {M.~I.} \bibnamefont
  {Katsnelson}},  and \bibinfo {author} {\bibfnamefont {A.~K.} \bibnamefont
  {Geim}}, }\Doi {10.1038/nphys1420} {\bibfield  {journal} {\bibinfo  {journal}
  {Nat. Phys.}\ }\textbf {\bibinfo {volume} {6}}, \bibinfo {pages} {30}
  (\bibinfo {year} {2010}{\natexlab{a}})}\bibfield  {note} {; preprint at
  \Eprint {http://arxiv.org/abs/arXiv:0909.1787} {\bibinfo {eprint}
  {arXiv:0909.1787}}}\BibitemShut {NoStop}%
\bibitem [{\citenamefont {Guinea} \emph
  {et~al.}(2010){\natexlab{b}}\citenamefont {Guinea}, \citenamefont {Geim},
  \citenamefont {Katsnelson}, and \citenamefont {Novoselov}}]{Guinea2010b}
  \BibitemOpen
  \bibfield  {author} {\bibinfo {author} {\bibfnamefont {F.}~\bibnamefont
  {Guinea}}, \bibinfo {author} {\bibfnamefont {A.~K.} \bibnamefont {Geim}},
  \bibinfo {author} {\bibfnamefont {M.~I.} \bibnamefont {Katsnelson}},  and
  \bibinfo {author} {\bibfnamefont {K.~S.} \bibnamefont {Novoselov}}, }\Doi
  {10.1103/PhysRevB.81.035408} {\bibfield  {journal} {\bibinfo  {journal}
  {Phys. Rev. B}\ }\textbf {\bibinfo {volume} {81}}, \bibinfo {pages} {035408}
  (\bibinfo {year} {2010}{\natexlab{b}})}\bibfield  {note} {; preprint at
  \Eprint {http://arxiv.org/abs/arXiv:0910.5935} {\bibinfo {eprint}
  {arXiv:0910.5935}}}\BibitemShut {NoStop}%
\bibitem [{\citenamefont {Min} \emph {et~al.}(2006)\citenamefont {Min},
  \citenamefont {Hill}, \citenamefont {Sinitsyn}, \citenamefont {Sahu},
  \citenamefont {Kleinman}, and \citenamefont {MacDonald}}]{Min2006}
  \BibitemOpen
  \bibfield  {author} {\bibinfo {author} {\bibfnamefont {H.}~\bibnamefont
  {Min}}, \bibinfo {author} {\bibfnamefont {J.~E.} \bibnamefont {Hill}},
  \bibinfo {author} {\bibfnamefont {N.~A.} \bibnamefont {Sinitsyn}}, \bibinfo
  {author} {\bibfnamefont {B.~R.} \bibnamefont {Sahu}}, \bibinfo {author}
  {\bibfnamefont {L.}~\bibnamefont {Kleinman}},  and \bibinfo {author}
  {\bibfnamefont {A.~H.} \bibnamefont {MacDonald}}, }\Doi
  {10.1103/PhysRevB.74.165310} {\bibfield  {journal} {\bibinfo  {journal}
  {Phys. Rev. B}\ }\textbf {\bibinfo {volume} {74}}, \bibinfo {eid} {165310}
  (\bibinfo {year} {2006})}\bibfield  {note} {; preprint at \Eprint
  {http://arxiv.org/abs/arXiv:cond-mat/0606504} {\bibinfo {eprint}
  {arXiv:cond-mat/0606504}}}\BibitemShut {NoStop}%
\bibitem [{\citenamefont {Gmitra} \emph {et~al.}(2009)\citenamefont {Gmitra},
  \citenamefont {Konschuh}, \citenamefont {Ertler}, \citenamefont
  {Ambrosch-Draxl}, and \citenamefont {Fabian}}]{Gmitra2009}
  \BibitemOpen
  \bibfield  {author} {\bibinfo {author} {\bibfnamefont {M.}~\bibnamefont
  {Gmitra}}, \bibinfo {author} {\bibfnamefont {S.}~\bibnamefont {Konschuh}},
  \bibinfo {author} {\bibfnamefont {C.}~\bibnamefont {Ertler}}, \bibinfo
  {author} {\bibfnamefont {C.}~\bibnamefont {Ambrosch-Draxl}},  and \bibinfo
  {author} {\bibfnamefont {J.}~\bibnamefont {Fabian}}, }\Doi
  {10.1103/PhysRevB.80.235431} {\bibfield  {journal} {\bibinfo  {journal}
  {Phys. Rev. B}\ }\textbf {\bibinfo {volume} {80}}, \bibinfo {pages} {235431}
  (\bibinfo {year} {2009})}\bibfield  {note} {; preprint at \Eprint
  {http://arxiv.org/abs/arXiv:0904.3315} {\bibinfo {eprint}
  {arXiv:0904.3315}}}\BibitemShut {NoStop}%
\bibitem [{\citenamefont {Guinea}(2010)}]{Guinea2010c}
  \BibitemOpen
  \bibfield  {author} {\bibinfo {author} {\bibfnamefont {F.}~\bibnamefont
  {Guinea}}, }\Doi {10.1088/1367-2630/12/8/083063} {\bibfield  {journal}
  {\bibinfo  {journal} {New J. Phys.}\ }\textbf {\bibinfo {volume} {12}},
  \bibinfo {pages} {083063} (\bibinfo {year} {2010})}\bibfield  {note} {;
  preprint at \Eprint {http://arxiv.org/abs/arXiv:1003.1618} {\bibinfo {eprint}
  {arXiv:1003.1618}}}\BibitemShut {NoStop}%
\bibitem [{\citenamefont {McCann}(2006)}]{McCann2006b}
  \BibitemOpen
  \bibfield  {author} {\bibinfo {author} {\bibfnamefont {E.}~\bibnamefont
  {McCann}}, }\Doi {10.1103/PhysRevB.74.161403} {\bibfield  {journal} {\bibinfo
   {journal} {Phys. Rev. B}\ }\textbf {\bibinfo {volume} {74}}, \bibinfo {eid}
  {161403} (\bibinfo {year} {2006})}\bibfield  {note} {; preprint at \Eprint
  {http://arxiv.org/abs/arXiv:cond-mat/0608221} {\bibinfo {eprint}
  {arXiv:cond-mat/0608221}}}\BibitemShut {NoStop}%
\bibitem [{\citenamefont {Enderlein} \emph {et~al.}(2010)\citenamefont
  {Enderlein}, \citenamefont {Kim}, \citenamefont {Bostwick}, \citenamefont
  {Rotenberg}, and \citenamefont {Horn}}]{Enderlein2010}
  \BibitemOpen
  \bibfield  {author} {\bibinfo {author} {\bibfnamefont {C.}~\bibnamefont
  {Enderlein}}, \bibinfo {author} {\bibfnamefont {Y.~S.} \bibnamefont {Kim}},
  \bibinfo {author} {\bibfnamefont {A.}~\bibnamefont {Bostwick}}, \bibinfo
  {author} {\bibfnamefont {E.}~\bibnamefont {Rotenberg}},  and \bibinfo
  {author} {\bibfnamefont {K.}~\bibnamefont {Horn}}, }\Doi
  {10.1088/1367-2630/12/3/033014} {\bibfield  {journal} {\bibinfo  {journal}
  {New J. Phys.}\ }\textbf {\bibinfo {volume} {12}}, \bibinfo {pages} {033014}
  (\bibinfo {year} {2010})}\BibitemShut {NoStop}%
\bibitem [{\citenamefont {Brey} and \citenamefont {Fertig}(2006)}]{Brey2006b}
  \BibitemOpen
  \bibfield  {author} {\bibinfo {author} {\bibfnamefont {L.}~\bibnamefont
  {Brey}} and \bibinfo {author} {\bibfnamefont {H.~A.} \bibnamefont {Fertig}},
  }\Doi {10.1103/PhysRevB.73.235411} {\bibfield  {journal} {\bibinfo  {journal}
  {Phys. Rev. B}\ }\textbf {\bibinfo {volume} {73}}, \bibinfo {pages} {235411}
  (\bibinfo {year} {2006})}\bibfield  {note} {; preprint at \Eprint
  {http://arxiv.org/abs/arXiv:cond-mat/0603107} {\bibinfo {eprint}
  {arXiv:cond-mat/0603107}}}\BibitemShut {NoStop}%
\bibitem [{\citenamefont {Han} \emph {et~al.}(2007)\citenamefont {Han},
  \citenamefont {{\"O}zyilmaz}, \citenamefont {Zhang}, and \citenamefont
  {Kim}}]{Han2007}
  \BibitemOpen
  \bibfield  {author} {\bibinfo {author} {\bibfnamefont {M.~Y.} \bibnamefont
  {Han}}, \bibinfo {author} {\bibfnamefont {B.}~\bibnamefont {{\"O}zyilmaz}},
  \bibinfo {author} {\bibfnamefont {Y.}~\bibnamefont {Zhang}},  and \bibinfo
  {author} {\bibfnamefont {P.}~\bibnamefont {Kim}}, }\Doi
  {10.1103/PhysRevLett.98.206805} {\bibfield  {journal} {\bibinfo  {journal}
  {Phys. Rev. Lett.}\ }\textbf {\bibinfo {volume} {98}}, \bibinfo {pages}
  {206805} (\bibinfo {year} {2007})}\bibfield  {note} {; preprint at \Eprint
  {http://arxiv.org/abs/arXiv:cond-mat/0702511} {\bibinfo {eprint}
  {arXiv:cond-mat/0702511}}}\BibitemShut {NoStop}%
\bibitem [{\citenamefont {Guo} \emph {et~al.}(2008)\citenamefont {Guo},
  \citenamefont {Guo}, and \citenamefont {Chen}}]{Guo2008}
  \BibitemOpen
  \bibfield  {author} {\bibinfo {author} {\bibfnamefont {Y.}~\bibnamefont
  {Guo}}, \bibinfo {author} {\bibfnamefont {W.}~\bibnamefont {Guo}},  and
  \bibinfo {author} {\bibfnamefont {C.}~\bibnamefont {Chen}}, }\Doi
  {10.1063/1.2943414} {\bibfield  {journal} {\bibinfo  {journal} {Appl. Phys.
  Lett.}\ }\textbf {\bibinfo {volume} {92}}, \bibinfo {eid} {243101} (\bibinfo
  {year} {2008})}\BibitemShut {NoStop}%
\end{thebibliography}
\end{document}